\documentclass[a4paper,11pt,notitlepage]{article}
\usepackage[notcite,notref]{}
\usepackage{latexsym}
\usepackage{amssymb}
\usepackage{color}
\usepackage{graphicx}
\usepackage{amsmath}
\usepackage{amsthm}
\usepackage{mathrsfs}
\usepackage{natbib}
\usepackage{amsfonts}
\usepackage{accents}
\usepackage{enumitem}
\usepackage{stmaryrd}     
\usepackage{subeqnarray}
\usepackage{subfigure}
\usepackage{float} 
\usepackage{bbm}
\usepackage{hyperref}

\newtheorem{theorem}{Theorem}[section]

\newtheorem{lemma}[theorem]{Lemma}
\newtheorem{remark}[theorem]{Remark}
\newtheorem{assumption}[theorem]{Assumption}

\newtheorem{prop}[theorem]{Proposition}

\newcommand{\al}{\alpha}
\newcommand{\alt}{\al_{t}}

\newcommand{\atil}{\tilde{\al}}
\newcommand{\CM}{\hat{\al}^{\vartheta}}

\newcommand{\mB}{b_0}

\newcommand{\XalT}{X_{T}^{\al}}
\newcommand{\Xalt}{X_{t}^{\al}}

\newcommand{\vtxbgl}{v^{\lambda, \gamma}(t,x,b)}

\newcommand{\Xitil}{\tilde{\chi}}
\newcommand{\Ktil}{\tilde{K}}
\newcommand{\Gamtil}{\tilde{\Gamma}}
\def \Rtil{\tilde{R}}

\newcommand{\dt}{\partial_t}
\newcommand{\dx}{\partial_x}

\newcommand{\gradb}{\nabla_b}
\newcommand{\grady}{\nabla_y}

\newcommand{\dxx}{\partial^{2}_{xx}}
\newcommand{\dxb}{\partial^{2}_{x b}}
\newcommand{\Hessb}{\mathcal{D}^2_b}

\newcommand{\trp}{\intercal}

\def\N{\mathbb N}
\def\P{\mathbb P}

\def\E{\mathbb E}
\def\P{\mathbb P}
\def\F{\mathbb F}
\def\R{\mathbb R}
\def\Rn{\mathbb R^n}

\newcommand{\Lnorm}{\left|}
\newcommand{\Rnorm}{\right|}

\usepackage[a4paper,top=3cm,bottom=3cm,left=3cm,right=3cm]{geometry}

\def \ep{\hbox{ }\hfill$\Box$}
\def \epR{\hbox{ }\hfill$\lozenge$}

\def\reff#1{{\rm(\ref{#1})}}

\def\beqs{\begin{eqnarray*}}
\def\enqs{\end{eqnarray*}}
\def\beq{\begin{eqnarray}}
\def\enq{\end{eqnarray}}

\title{Bayesian learning for the Markowitz portfolio selection problem \footnote{This work is issued from a CIFRE collaboration between OSSIAM and LPSM.}}
\author{Carmine De Franco\footnote{OSSIAM, E-mail:
    carmine.de-franco@ossiam.com}\and Johann Nicolle\footnote{OSSIAM and LPSM, Universit\'e Paris Diderot, E-mail:
    johann.nicolle@ossiam.com}\and Huy\^en Pham \footnote{LPSM, Universit\'e Paris Diderot, E-mail: pham@lpsm.paris}}
    
\begin{document}

\frenchspacing

\maketitle

\begin{abstract}
We study  the Markowitz portfolio selection problem with unknown drift vector in the multidimensional framework. The prior belief on the uncertain expected rate of return is modeled by an arbitrary  probability law, and a Bayesian approach from filtering theory is used to learn the posterior distribution about the drift given the observed market data of the assets. The Bayesian Markowitz problem is then embedded into an auxiliary standard control problem that we characterize by a dynamic programming method and prove the existence and uniqueness of a smooth solution to the related semi-linear partial differential equation (PDE).  The optimal Markowitz portfolio strategy is explicitly computed in the case of a Gaussian prior distribution. Finally, we measure the quantitative impact of learning, updating the strategy from observed data, compared to non-learning, using a constant drift in an uncertain context, and analyze the sensitivity of the value of information w.r.t. various relevant parameters of our model. 
\end{abstract}

\medskip


\medskip

\numberwithin{equation}{section}

\section{Introduction}

Portfolio selection is a core problem in mathematical finance and investment management. Its purpose is to choose the best portfolio according to a criterion of optimality. The mean-variance optimization provides a criterion of optimality considering the best portfolio as the one that maximizes an expected level of return given a certain level of risk, the one that the investor can bear, or conversely, minimizes the risk of a portfolio given an expected level of return. \cite*{Markowitz1952} pioneered modern portfolio theory by settling the basic concepts. This conceptual framework, still widely used in the industry, leads to the efficient frontier principle which exhibits an intuitive relationship between risk and return. 

Later, portfolio selection theory was extended several times to encompass multi-period problems, in discrete time by \cite*{Samuelson1969} and in continuous time by \cite*{Merton1969}. 
 \cite{Karatzas1987} made a decisive step forward when they solved Merton's problem for a large class of smooth utility functions using dual martingale methods under the no-bankruptcy constraint.

Originally, the literature on portfolio selection assumed that the parameters of the model, drifts and volatilities, are known and constant. This assumption raises the question of estimating these parameters. 
Typically, the parameters are estimated from past data and are fixed once and for all. This convenient solution does not look realistic in practice since it does not adapt to changing market conditions. Moreover, as \cite*{Merton1980} among others showed, estimates of variances and covariances of the assets are more accurate than the estimates of the means. Indeed, he demonstrated the slow convergence of his estimators of the instantaneous expected return in a log-normal diffusion price model. Later, \cite*{Best1991} argued that mean-variance optimal portfolios are very sensitive to the level of expected returns. 
As a consequence, estimating the expected return is not only more complicated than estimating the variance/covariance, but also a wrong estimation of the expected return can result in a very suboptimal portfolio a posteriori.  

To circumvent these issues, an extensive literature incorporating parameters uncertainty in portfolio analysis has emerged, see for example \cite*{Barry1974} and \cite*{Klein1976} and methods using Bayesian statistics have been developed, see  \cite*{Frost1986}, \cite*{Aguilar2000}, \cite*{Avramov2010} and \cite{Bodnar2017}.  
In particular, the \cite*{Black1992} model, based on economic arguments and equilibrium relations, provides more stable and diversified portfolios than simple mean-variance optimization. 
Based upon the Markowitz problem and the capital asset pricing model (CAPM), this model remains static and cannot benefit from the flow of information fed by the market prices of the assets. 
This loss of information is detrimental in the optimization process, since it does not allow for an update of the model given the most recent available information.

Consequently, research then focused on taking advantage of the latest information conveyed by the prices. It explains the subsequent growing literature on filtering and learning techniques in a partial information framework, see \cite*{Lakner1995}, \cite*{Lakner1998}, \cite*{Rogers2001}  and \cite{Cvitanic2006}. 
The most noticeable research involving optimization and Bayesian learning techniques is by \cite*{Karatzas2001}. Using martingale methods, they computed the optimal portfolio allocation for a large class of utility functions in the case of an unknown drift and Gaussian asset returns. 
\cite*{Gueant2017} extend the previous results to take into account both the liquidity and the expected returns of the assets, coupling Bayesian learning and dynamic programming techniques while addressing optimal portfolio liquidation and transition problems. Recently, \cite{Bauder2018a} suggest to deal with the Markowitz problem using the posterior predictive distribution. 
A Bayesian efficient frontier is also derived and proved to outperform the overoptimistic sample efficient frontier.   

In this paper, we consider an investor who is willing to invest in an optimal portfolio in the Markowitz sense for a finite given time horizon. 
The investment universe consists of assets for which we assume the covariance matrix known and constant. The drift vector is uncertain and assumed to have a prior probability distribution. 

We contribute to the literature of optimal portfolio selection by formulating and solving, in the multidimensional case, the Markowitz problem in the case of an uncertain drift modeled by a prior distribution. This a priori time inconsistent problem cannot be tackled by convex duality and martingale method as in \cite*{Karatzas2001}. 
Instead, we adapt the methodology of \cite*{Zhou2000} to our Bayesian learning framework, and show how the Bayesian-Markowitz problem can be embedded into an auxiliary standard control problem that we study by the dynamic programming approach.
Optimal strategies are then characterized in terms of a smooth solution to a semi-linear partial differential equation, and we highlight the effect of Bayesian learning compared to classical strategies based on constant drift. 
Although prior conjugates are widely used in the literature, here we extend to the multidimensional case, a recent result by \cite*{Ekstrom2016} which enables us to use any prior distribution with invertible covariance matrix to model the uncertain drift allowing for a wide range of investors' strategies.
In the particular case of a multidimensional Gaussian prior conjugate, we are able to exhibit a closed-form analytical formula. 

Next, we measure the benefit of learning on the Markowitz strategy. 
To do so, we compare the Bayesian learning strategy to the subsequently called non-learning strategy which considers the coefficients of the model, especially the drift, constant. 
The Bayesian learning strategy is characterized by the fact that it uses, as information, the updated market prices of the assets to adjust the investment strategy and reaches optimality in the Markowitz sense. 
On the other hand, the non-learning strategy is a typical constant parameter strategy that estimates once and for all its unknown parameters with past data at time $0$, leading to sub-optimal solutions to the Markowitz problem because it misses the most recent information.
The benefit of learning, measured as the difference between the Sharpe ratios of the portfolios based on the Bayesian learning and the non-learning strategies, is called the value of information in the sequel. 
We then analyze in a one-dimensional model the sensitivities of the value of information w.r.t. four essential variables: drift volatility, Sharpe ratio of the asset, time and investment horizon. 

The paper is organized as follows:
Section \ref{sec: Markowitz_pb} depicts the Markowitz problem in our framework of drift uncertainty modeled by a prior distribution and the steps to turn it into a solvable problem.
Section \ref{sec: BL} introduces the Bayesian learning framework and the methodology to consider any prior with positive definite covariance matrices. Section \ref{sec: Sol_to_BM_prob} provides the main results and some specific examples where some computations are possible analytically.
Section \ref{sec: Impact_Learning} concerns the value of information and its sensitivity to different parameters of an unidimensional model: the volatility of the drift, the sharpe ratio of the asset, the time, and the investment horizon.
Section \ref{sec: Conclusion} concludes this paper. 

\section{Markowitz problem with prior law on the uncertain drift} \label{sec: Markowitz_pb}

We consider a financial market on a probability space ($\Omega$,$\mathcal{F}$,$\P$) equipped with a filtration $(\mathcal{F}_t)_{t \geq 0}$ that satisfies the usual conditions, and on which is defined a standard  $n$-dimensional Brownian motion $W$. The market consists of $n$ risky assets that are continuously traded. The assets prices process $S$  is defined as: 
  \beq \label{eq: model_unc_drift}
    \left \{
    \begin{array}{rcl}
      dS_t & = &\text{diag}(S_t) \big( Bdt + \sigma dW_t \big ), \quad t \in [0,T],\\
      S_0 & = & s_0 \in\Rn,
    \end{array}\right.
  \enq
where $T$ is the investment horizon, $B$ is a random vector in $\Rn$, independent of $W$, with distribution $\mu$ and such that $\E[|B|^2] < \infty$. The prior law of $B$ represents the subjective beliefs of the investor about the likelihood of the different values that $B$ might take.  
  \\
  
  In the sequel, we consider the following assumptions satisfied:
\begin{assumption}
	The covariance matrix ${\rm Cov}\left(B\right)$ of the random vector $B$ is positive definite, \label{ass: B_pos}
\end{assumption}
\begin{assumption}
    The $n \times n$ volatility matrix $\sigma$ is known and invertible, and we denote $\Sigma$ $=$ $\sigma\sigma^\trp$. \label{ass: sig}
\end{assumption}
  
  We write $\mathbb{F^S}= \{ \mathcal{F}^S_t \}_{t \geq 0}$ the filtration generated by the price process $S$ augmented by the null sets of $\mathcal{F}$. The filtration $\mathbb{F}^S$ represents the only available source of information: the investor  observes the price process $S$ but not the drift. 

  We denote by $\al = (\al_t)_{0\leq t \leq T}$ an admissible investment strategy representing the amount invested in each of the $n$ risky assets. It is a $n$-dimensional 
  $\mathbb{F^S}$-progressively measurable process, valued in $\R^{n}$, and  satisfying the following integrability condition
  \beqs
  	\E \bigg[\int_0^\trp |\al_t|^2dt\bigg]  < \infty.
  \enqs
  We define $\mathcal{A}$ as the set of all admissible investment strategies. 

  The evolution of the self-financing wealth process $X^{\al}$, given $\al \in \mathcal{A}$ and an initial wealth $x_{0} \in\R$, is given by:
  \beq\label{eq: process_X}
  	d\Xalt = \alt^\trp \text{diag}(S_t)^{-1}dS_t =\alt^\trp (B dt + \sigma dW_{t}), \quad 0 \leq t \leq T, \quad X^{\al}_0 = x_{0}.
  \enq

  The Markowitz problem is then formulated as :
\beq   \label{eq: markowitz_problem}
 U_{0}(\vartheta) &:=& \sup_{\alpha\in\mathcal{A}} \Big\{  \E[\XalT] :   {\rm Var}(\XalT) \leq \vartheta \Big\}, 
\enq
where $\vartheta>0$ is the variance budget the investor can allow on her portfolio. The value function $U_0$ is then the expected optimal terminal wealth value the investor can achieve given her variance constraint. It is very standard to transform the initial problem \eqref{eq: markowitz_problem} into a so called mean-variance problem, where the constraint becomes part of the objective function:
\beq \label{eq: mean_variance_problem}
    	V_0(\lambda) \; := \;  \inf_{\al\in\mathcal A} \big[ \lambda {\rm Var}(\XalT) - \E[\XalT] \big], \;\;\; \lambda > 0. 
\enq

The well-known connection between the Markowitz and mean-variance problem is stated in the following lemma.

\begin{lemma} \label{lem: dual_transform}
We have the duality relation
 \beqs \label{eq: dual transform}
   \left\{
    \begin{array}{rcl}
     V_{0}(\lambda) & = &  \inf_{\vartheta>0}\big[\lambda \vartheta - U_{0}(\vartheta)\big],\;\;\;  \forall \lambda>0  \\
     U_{0}(\vartheta ) & = &  \inf_{\lambda>0}\big[\lambda \vartheta - V_{0}(\lambda)\big], \;\;\;  \forall\vartheta>0.  
    \end{array}
    \right.
    \enqs
Furthermore, if $\al^{*, \lambda}$ is an optimal control for problem \eqref{eq: mean_variance_problem}, then 
$\hat{\al}^{\vartheta}$ $:=$ $\al^{*, \lambda(\vartheta)}$ is an optimal control for problem $U_0(\vartheta)$ in \eqref{eq: markowitz_problem}, with
$\lambda(\vartheta)$ $:=$ ${\rm arg}\min_{\lambda>0}\big[\lambda \vartheta - V_{0}(\lambda)\big]$, and ${\rm Var}(X_T^{\hat\alpha^{\vartheta}})$ $=$ $\vartheta$.   
 \end{lemma}

The proof of this result is standard and detailed in \ref{sub:proof_lemma_duality}.
\\

Because of the variance term, mean-variance problem \eqref{eq: mean_variance_problem} does not fit into a classical time consistent control problem. 
To circumvent this issue, we adopt a similar approach as in \cite*{Zhou2000} in order to embed it into an auxiliary control problem, for which we can use the dynamic programming method. 
 
\vspace{1mm}
 
  \begin{lemma} \label{lem: quadratic_transform}
    The mean-variance optimization problem \eqref{eq: mean_variance_problem} can be written as follows:
    \beq \label{eq: V1_quadratic}
    	V_0(\lambda) \; = \;   \inf_{\gamma\in\R} \big[ \tilde{V_0}\left(\lambda, \gamma\right) + \lambda\gamma^2\big]
    \enq
    where
    \beq \label{eq: V1_quadratic_def}
    \tilde{V_0}(\lambda, \gamma) \; := \;  \inf_{\al \in\mathcal A}\big[ \lambda \E[(X^\al_T)^{2}]-(1+2\lambda\gamma) \E[X^\al_T]\big]. 
    \enq
  Moreover, if $\atil^{\lambda, \gamma}$ is an optimal control for \eqref{eq: V1_quadratic_def} and $\gamma^*(\lambda)$ $:=$ 
  ${\rm argmin}_{\gamma\in\R} \big[\tilde V_0(\lambda, \gamma) +\lambda\gamma^2\big]$ then $\al^{*, \lambda}$ $:=$ 
  $\atil^{\lambda, \gamma^*(\lambda)}$ is an optimal control for $V_0(\lambda)$ in  \eqref{eq: mean_variance_problem}.
  \\
  
  \noindent Furthermore, the following equality holds true
    \beq\label{eq: opt_gamma}
    	\gamma^*(\lambda) 
	\; = \;  \E[X^{\alpha^{*, \lambda}}_T].
    \enq
  \end{lemma}

\vspace{1mm}

 The proof of this lemma follows \cite*{Zhou2000} and is provided in \ref{sub:proof_lemma_quadratic_transform}. 
 As we shall see in Section  \ref{sec: Sol_to_BM_prob},  problem  \reff{eq: V1_quadratic_def} is well suited for dynamic programming. 

 \vspace{2mm}
 
We end this section by recalling the solution to the Markowitz problem in the case of constant known drifts and volatilities, which will serve as a benchmark  to refer to.

  \begin{remark}[Case of constant known drift] \label{rem: constant_parameter}
  {\rm 
    In the case where the drift vector $b_0$ and the volatility matrix $\sigma$ are known and constant, we have the following results :
    \begin{itemize}
      \item The value function of the mean-variance problem with initial wealth $x_0$ is given by
      \beqs \label{eq: value_function_BS}
      	V_0(\lambda) \;= \;  v^\lambda(0,x_0) = -\frac{1}{4\lambda}\left( e^{|\sigma^{-1}b_0|^2T} -1\right)-x_0
      \enqs
      with
      \beqs \label{eq: V_0_BS}
      	v^\lambda(t,x) &=&  \lambda e^{-|\sigma^{-1}b_0|^2(T-t)}\left(x-x_0\right)^{2}-\frac{1}{4\lambda}\left( e^{|\sigma^{-1}b_0|^2(T-t)} -1\right)-x.
      \enqs
      The corresponding optimal portfolio  strategy for $V_0(\lambda)$ is  given in feedback form  by 
      \beqs
      	\alpha_t^{*,\lambda}  \; = \; a_0^{\lambda}(t, X^*_t),
      \enqs
      where
      \beqs 
      	a_0^{\lambda}(t,x) &:=& \Big( x_0-x + \frac{e^{|\sigma^{-1}b_0|^2T}}{2 \lambda}\Big)\Sigma^{-1}b_0 
      \enqs
      and $X^*_t$ is the optimal wealth obtained with the optimal feedback strategy.
      \item The optimal portfolio  strategy of the Markowitz problem $U_0(\vartheta)$ is  then given by 
      \beqs
      	\CM_t  \; = \; \alpha_t^{*,\lambda_0(\vartheta)} 
        \enqs
      where
      \beqs
       \lambda_0(\vartheta) &:=& \sqrt{\frac{e^{|\sigma^{-1}b_0|^2T}-1}{4 \vartheta}}.     
      \enqs
 Moreover, the solution to the Markowitz problem is
       \beqs
      U_0(\vartheta) &=& x_0 + \sqrt{\vartheta\big( e^{|\sigma^{-1}b_0|^2T}-1 \big)}. 
      \enqs
    \end{itemize}
See for instance \cite*{Zhou2000} for more details.
 }   
 \epR
  \end{remark}

\section{Bayesian learning}\label{sec: BL}

Since the investor does not observe the assets drift vector $B$, she needs to have a subjective belief on its potential value. It is represented as a prior probability distribution $\mu$ on $\R^n$, assumed to satisfy for some $a>0$, 
\beqs
	\int_{\Rn} e^{a|b|^2}\mu(db) < \infty .
\enqs
The prior probability distribution will then learn and infer the value of the drift from observable samples of the assets prices. Using filtering theory, and following in particular the recent work by \cite*{Ekstrom2016} that we extend to the multi-dimensional case, we compute the posterior probability distribution of $B$ given the observed assets prices.  

\vspace{1mm}

Let us first introduce  the  $\R^n$-valued  process $Y_t = \sigma^{-1}Bt + W_t$, which clearly generates the same filtrations as $S$, and write the observation process
\beqs
S_t^i &=& S_0^i \exp\Big( \sigma^i Y_t  - \frac{|\sigma^i|^2}{2} t \Big), \;\;\;  i =1,\ldots,n, \;\; 0 \leq t \leq T, 
\enqs
where $\sigma^i$ denotes the $i$-th line of the matrix $\sigma$. 

Equivalently, we can express $Y$ in terms of $S$ as: $Y_t$ $=$ $h(t,S_t)$ where $h$ $:$ $[0,T]\times (0,\infty)^n$ $\rightarrow$ $\R^n$ is defined by
\beqs
h(t,s) &:=& \sigma^{-1} \left(
\begin{array}{c}
\ln\big(\frac{s^1}{S_0^i}\big) + \frac{|\sigma^1|^2}{2} t \\
\vdots \\
\ln\big(\frac{s^n}{S_0^i}\big) + \frac{|\sigma^n|^2}{2} t 
\end{array}
\right), \;\;\;  s = (s^1,\ldots,s^n) \in (0,\infty)^n. 
\enqs

The following result gives the conditional distribution of $B$ given observations of the assets market prices in terms of the current value of $Y$. We refer to Proposition 3.16 in \cite*{Bain2009} as well as \cite*{Karatzas2001} and \cite*{Pham2008} for a proof. 

\begin{lemma}\label{lem: Ekstrom multidim}
Let  $g: \Rn \to \Rn$ satisfying $\int_{\Rn}|g(b)| \mu(db) < \infty$. Then
\beqs
\E\big[ g(B) | {\cal F}_t^S \big] \; = \; \E[g(B)|Y_t]   &=&  
\frac{\int_{\Rn}g(b)e^{<\sigma^{-1}b,Y_t>-\frac{| \sigma^{-1}b |^2}{2}t}\mu(db)}{\int_{\Rn}e^{<\sigma^{-1}b,Y_t>-\frac{| \sigma^{-1}b |^2}{2}t}\mu(db)}.
 \enqs
\end{lemma}

\vspace{5mm}

From Lemma \ref{lem: Ekstrom multidim},  the conditional distribution $\mu_{t,y}$ of $B$ given $Y_t=y$ is given by
  \beq \label{eq: mu(t,y)}
  \mu_{t,y}(db)  &=&  \frac{e^{<\sigma^{-1}b,y>-\frac{| \sigma^{-1}b |^2}{2}t}\mu(db)}{\int_{\Rn}e^{<\sigma^{-1}b,y>-\frac{| \sigma^{-1}b |^2}{2}t}\mu(db)},
  \enq
and the posterior predictive mean of the drift is
\beqs
\hat{B}_t \; := \;  \E[B|\mathcal{F}_t^S] \; = \;  \E[B|Y_t] &=&  f(t,Y_t), 
\enqs
where 
\beq \label{eq: f(t,y)}
  f(t,y) &:=&  \int_{\Rn}b \mu_{t,y}(db) \;= \;  \frac{\int_{\Rn}be^{<\sigma^{-1}b,y>-\frac{| \sigma^{-1}b |^2}{2}t}\mu(db)}{\int_{\Rn}e^{<\sigma^{-1}b,y>-
  \frac{| \sigma^{-1}b |^2}{2}t}\mu(db)}.
  \enq

  The following result shows some useful properties regarding the function $f$.

  \begin{lemma}\label{lem: function_f}
  For each $t\in [0, T]$, the function $f_t$ $:$ $\R^n\to\R^n$ defined as $f_t(y)$ $:=$ $f(t, y)$, where $f$ is given in \eqref{eq: f(t,y)} is invertible when restricted to its image 
  ${\cal B}_t$ $:=$ $f_t(\R^n)$, and we have  
  \beqs
      \nabla_y f(t, y)  \;= \;  {\rm Cov}\left(B\left|Y_t = y \right.\right)\left(\sigma^{-1}\right)^\trp, \;\;\; y \in \R^n. 
    \enqs
 and the conditional covariance matrix of $B$ given $Y_t$ $=$ $y$ is  positive definite. 
 \end{lemma}
\noindent {\bf Proof.} 
The vector-valued function $y \mapsto f_t(y)$ is differentiable, hence we can compute the matrix function $\nabla_y f$ element-wise,
	 \begin{align*}
      \left[ \partial_{y_j} f_t(y)  \right]_i = & \int_{\Rn} [\sigma^{-1}b]_j b_i\mu_{t,y}(db)
      -\int_{\Rn} b_i \mu_{t,y}(db)\int_{\Rn} [\sigma^{-1}b]_j \mu_{t,y}(db) \\ 
      = & \sum_{k=1}^{n} \left( \sigma^{-1}_{j,k} \int_{\Rn} b_k b_i\mu_{t,y}(db)-\int_{\Rn} b_i \mu_{t,y}(db)\int_{\Rn} b_k \mu_{t,y}(db) \right)\\
      = & \sum_{k=1}^{n} \sigma^{-1}_{j,k}  \text{cov}(B_i,B_k|Y_t=y)\\
      = & \left( \sigma^{-1} \text{cov}\left(B\left|\right.Y_t=y\right) \right)_{j,i}\\
      = & \left(\text{cov}\left(B\left|\right.Y_t=y\right)(\sigma^{-1})^\trp \right)_{i,j}.
    \end{align*}
    
    If $\text{cov}\left(B\left|\right.Y_t\right)$ is not positive definite, then for some linear combination of $B$, we would have $\text{Var}\left(\sum_j q_j B_j\left.\right| Y_t\right) = 0$ for some $q\in\Rn$, meaning that $\sum_j q_j B_j = C,\, C\in\R,\, \mu_{t,y}$-a.e.. But from \eqref{eq: mu(t,y)}, this would imply $\sum_j q_j B_j = C,\, C\in\R,\, \mu$-a.e which contradicts Assumption \ref{ass: B_pos}.
\\

	Fix $t \geq 0$, the function $f_t$ $:$ $\Rn \to \Rn$ is obviously surjective on its image ${\cal B}_t$. To show that it is injective, we define $\tilde{f} = f_t\sigma^\trp$ and take two $\Rn$-vectors $y_1 \neq y_2$. If we compute the Taylor expansion of $\tilde{f}$,
    \beqs
		\tilde{f}(y_1)-\tilde{f}(y_2) = \grady \tilde{f}(\eta)(y_1-y_2),
	\enqs
	where $\eta$ is a convex combination of $y_1$ and $y_2$. Since $ \text{cov}(B|Y_t)$ is positive definite $\grady \tilde{f}$ is positive definite so $\tilde{f}(y_1) \neq\tilde{f}(y_2)$ and $f_t$ is injective.

\ep 
\\

Let us now introduce the so-called innovation process 
\beqs
\hat{W}_t &: =&  \sigma^{-1}\int^t_0(B-\hat{B}_s)ds+W_t, \;\;\; 0 \leq t \leq T,
\enqs
which is a $(\P,\F^S)$-Brownian motion, see Proposition 2.30 in \cite*{Bain2009}. The observation process $Y$ is written in terms of the innovation process as 
 
\beq \label{eq: alt_form_Y}
  	Y_t \;= \;  \sigma^{-1}\int_0^t \hat B_s ds + \hat W_t.
  \enq
Applying  It\^o's formula to $\hat{B}_t = f(t,Y_t)$ with \eqref{eq: alt_form_Y}, and recalling that  $\hat B$ is a $(\P,\F^S)$-martingale,  we see that  $d\hat{B}_t$ $=$  
$\nabla_y f(t,Y_t) d \hat{W}_t$. By Lemma \ref{lem: function_f}, and defining the matrix-valued function $\psi$ by  
  \beq\label{eq: psi}
  	\psi(t,b) \; := \; \nabla_y  f(t,f_t^{-1}(b)), \;\;\; t \in [0,T], \; b \in {\cal B}_t,
  \enq
  where $y = f_t^{-1}(b)$ is the unique value of the observation process $Y_t$ that yields $\hat{B}_t = b$, we have 
  \beq \label{eq: dhatB psi}
  	d\hat{B}_t &= &  \psi(t,\hat{B}_t)d\hat{W}_t.
  \enq

\begin{remark}[Dirac case]
The Dirac prior distribution $\mu = \mathbbm{1}_{b_0}$ does not verify Assumption \reff{ass: B_pos}. Indeed in this case, $ \forall$ $(t,b)$ $f(t,b):=b_0$, consequently we cannot properly define the function $f_t^{-1}$ and the matrix-valued function $\psi$. A natural way to extend our framework to the Dirac case is by setting $\forall (t,b),$ $\psi(t,b):=0$. Indeed when $B$ is known then $cov(B) = 0$ and thus $\psi=0$. 
\epR
\end{remark}

  We can rewrite Model \eqref{eq: model_unc_drift} in terms of observable variables:
  \beqs
    \left \{
    \begin{array}{rllc}
      dS_t & =   \text{diag}(S_t)\left( \hat B_t dt + \sigma d\hat W_t\right), & \;t\in [0,T],  & \; S_0 \; = \; s_0, \\
      d\hat B_t & = \psi(t,\hat{B}_t)d\hat{W}_t, &\; t \in [0,T],  & \;\hat B_0 \; = \; b_0  \; = \E[B]. \\
    \end{array}\right.\label{eq: model_crt_drift}
  \enqs
 Notice that the dynamics of the observable process $\hat B_t$ $=$ $f(t,Y_t)$ $=$ $f(t,h(t,S_t))$ is fully determined by the function $\psi$ given in analytical form by  \eqref{eq: psi} (explicit computations will be given later in the next sections).  The dynamics of the  wealth process in 
 \eqref{eq: process_X} becomes
  \beq\label{eq: process_X_hat}
  	d\Xalt \;=\; \alt^\trp \text{diag}(S_t)^{-1}dS_t =\alt^\trp (\hat B_t dt + \sigma d\hat W_{t}), \quad 0 \leq t \leq T, \quad X^{\al}_0 = x_{0}.
  \enq

\section{Solution to the Bayesian-Markowitz problem}\label{sec: Sol_to_BM_prob}

  \subsection{Main result}

 In order to solve the initial Markovitz problem in \eqref{eq: markowitz_problem}, we first solve problem \eqref{eq: V1_quadratic_def}, then use Lemma \ref{lem: quadratic_transform} to find the optimal $\gamma$ and obtain the dual function $V_0(\lambda)$ stated in \eqref{eq: V1_quadratic}, and finally apply Lemma \ref{lem: dual_transform} to find the optimal Lagrange multiplier $\lambda$ which gives us the original value function $U_0(\vartheta)$ and the associated optimal strategy. 
 
 \vspace{1mm}

 Let us define the dynamic value function associated to problem \eqref{eq: markowitz_problem}:  
 \beq
    v^{\lambda,\gamma}(t,x,b) := \inf_{\al \in \mathcal{A}} J^{\lambda,\gamma}(t,x,b,\al), \;\;\;\;\; t \in [0,T], \; x \in \R, \; b \in {\cal B}_t,  \label{eq: value function} 
  \enq
  with
  \beq
	J^{\lambda, \gamma}(t,x,b,\al) :=  \E \Big[ \lambda (X^{t,x,b,\al}_T)^2 -(1+2\lambda \gamma) X^{t,x,b,\al}_T\Big], \nonumber 
  \enq 
  where $X^{t,x,b,\al}$ is the solution to  \eqref{eq: process_X_hat} on $[t,T]$, starting at 
  $X^{t,x,b,\al}_t = x$  and $\hat B_t = b$ at time $t$ $\in$ $[0,T]$,  controlled by $\alpha$ $\in$ ${\cal A}$, so that 
$\tilde{V_0}(\lambda, \gamma)$ $=$  $v^{\lambda,\gamma}(0, x_0, b_0)$, with $b_0$ $=$ $\E[B]$. 
  
  \vspace{2mm}
  
 Problem \reff{eq: value function} is a standard stochastic control problem that can be characterized by the dynamic programming 
 Hamilton-Jacobi-Bellman (HJB) equation, which is a fully nonlinear PDE. 
 Actually, by exploiting the quadratic structure of this control problem (as detailed below), the HJB equation can be reduced to the following semi-linear PDE 
 on ${\cal R}$ $=$ $\{ (t,b): t \in [0,T], b \in {\cal B}_t\}$:  
 \beq \label{eq: PDE_w}
 	-\dt{R} - \frac{1}{2} \text{tr}\big( \psi \psi^\trp \Hessb R \big) +  \tilde{F}(t,b,\gradb R)= 0, 
  \enq
  with
  \beq \label{eq: DefFtild}
  		\tilde{F}(t,b,p) := 2 \left(\psi \sigma^{-1}b \right)^\trp p - \frac{1}{2} |\psi^\trp p|^2 -|\sigma^{-1}b|^2,
  \enq 
  and terminal condition
  \beq \label{eq: termcond}
  	 R(T,b) = 0, \quad b \in {\cal B}_t,
  \enq 

We can then state the main result of this paper, which provides the analytic solution to the Bayesian Markowitz problem. 

  \begin{theorem} \label{th: theomain}
    Suppose there exists a solution $R \in C^{1,2}([0,T) \times {\cal B}_t; \R_+) \cap C^0({\cal R}; \R_+)$ to the semi-linear Eq. \eqref{eq: PDE_w} with terminal condition \eqref{eq: termcond}. Then, for any $\lambda$ $>$ $0$, 
    \beqs
    	V_0(\lambda) \; = \;  -\frac{1}{4 \lambda} \left(e^{R(0,b_0)}-1 \right)-x_0
    \enqs
    with the associated optimal mean-variance control given in feedback form by 
    \beq \label{abayes}
\alpha^{*, \lambda}_t &=&    a_0^{Bayes,\lambda}( t, X_t^{\alpha^{*,\lambda}}, \hat{B}_t) \;= \;  a_0^{Bayes,\lambda}\big( t, X_t^{\alpha^{*,\lambda}}, f(t,h(t,S_t))\big)
    \enq
    where
    \beq \label{eq: a_bayes}
    a_0^{Bayes,\lambda}(t,x,b)  &:=&  \Big(x_0-x+\frac{e^{R(0,b_0)}}{2\lambda}\Big) \left( \Sigma^{-1} b - \left(\psi \sigma^{-1}\right)^\trp \gradb R(t,b) \right),
    \enq
and the corresponding optimal terminal wealth is equal to
\beq \label{eq: optterminal}
\E\big[X_T^{\alpha^{*,\lambda}} \big] &=& x_0  + \frac{1}{2\lambda} \big(e^{R(0,b_0)} -1\big).
\enq    
Moreover, for any $\vartheta$ $>$ $0$, 
    \beqs
    	U_0(\vartheta) \; = \;  x_0 + \sqrt{ \vartheta \left( e^{R(0,b_0)}-1\right)}
    \enqs
    with the associated optimal Bayesian-Markowitz strategy given by 
    \beqs
    	\CM \; = \;  \alpha^{*, \lambda(\vartheta)},
    \enqs
    with
    \beq \label{eq: explilambda}
    	\lambda(\vartheta) \; = \;  \sqrt{\frac{e^{R(0,b_0)}-1}{4 \vartheta}}.
    \enq	
  \end{theorem}
\noindent {\bf Proof.} The detailed proof is postponed in \ref{proof prop: th: main}, and we only sketch here the main arguments. 

For fixed $(\lambda,\gamma)$, the HJB equation associated to the standard stochastic control problem \reff{eq: value function}  is written as
    \begin{equation} \label{eq: HJB1}	
        \dt v^{\lambda,\gamma} + \frac{1}{2} \text{tr} \big( \psi\psi^\trp\Hessb v^{\lambda,\gamma} \big) + \inf_{\al \in \mathcal{A}} \Big[\al^\trp  b \dx v^{\lambda,\gamma} +\frac{1}{2}|\sigma^\trp \al|^2  \dxx v^{\lambda,\gamma} + \al^\trp \sigma \psi^\trp  \dxb v^{\lambda,\gamma}  
        \Big ] \; = \;  0,  
    \end{equation}
    with terminal condition    
    \beqs
        & v^{\lambda,\gamma}(T,x,b) \; = \;   \lambda x^{2} -(1+2\lambda \gamma) x.
	\enqs
We look for a solution in the ansatz form: $v^{\lambda,\gamma}(t,x,b)$ $=$ $K(t,b) x^2 + \Gamma(t,b) x + \chi(t,b)$. By plugging into the  above HJB equation, and identifying the terms in $x^2$, $x$, we find after some straightforward calculations that 
    \beqs
    	v^{\lambda,\gamma}(t,x,b) \;= \;  e^{-R(t,b)} \left[ \lambda x^2 - (1+2 \lambda \gamma)x + \frac{(1+2 \lambda \gamma)^2}{4 \lambda}\right] -\frac{(1+2 \lambda \gamma)^2}{4 \lambda},
    \enqs
 where $R$ has to satisfy  the semi-linear PDE    \reff{eq: PDE_w} (which does not depend on $\lambda,\gamma$). Since $R$ is assumed to exist smooth, the optimal feedback control achieving the argmin in the HJB equation \reff{eq: HJB1}  is given by
    \beq\label{eq: Markowitz controls}
    	\tilde \al_t ^{\lambda, \gamma}= \tilde a^{ \lambda, \gamma}(t, X^{\tilde \al ^{\lambda, \gamma}}_t,\hat B_t),
    \enq
    with
    \beq \label{eq: atild}
		\tilde a^{\lambda, \gamma}(t,x,b) \;:=\;  \Big(\frac{1}{2 \lambda}+\gamma-x\Big) \big( \Sigma^{-1} b - (\psi \sigma^{-1})^\trp \gradb R(t,b) \big),
	\enq
while $\gamma^*(\lambda)$ $:=$   $\text{arg}\min_{\gamma\in\R} \big[\tilde V_0(\lambda, \gamma) +\lambda\gamma^2\big]$ $=$ 
$\text{arg}\min_{\gamma\in\R} \big[\tilde v^{\lambda, \gamma}(0,x_0,b_0) +\lambda\gamma^2\big]$ is given by
   \beq \label{eq: gammalambda}
    	\gamma^*(\lambda)  & = &  x_0  + \frac{1}{2\lambda} \big(e^{R(0,b_0)} -1\big),
    \enq
    which leads to the expression \reff{eq: optterminal} from \reff{eq: opt_gamma}. 
We deduce that 
    \beq \label{eq: V_0_lambda}
         V_0(\lambda) &=&  \tilde{V}_0(\lambda, \gamma^*(\lambda)) + \lambda (\gamma^*(\lambda))^2 \nonumber \\
        & = & v^{\lambda, \gamma^*(\lambda)}(0,x_0,b_0) + \lambda (\gamma^*(\lambda))^2 \nonumber \\
        & =&  -\frac{1}{4 \lambda} \left(e^{R(0,b_0)}-1 \right)-x_0. \label{eq: V0lambda}
   \enq
    From Lemma \ref{lem: quadratic_transform}, we know $\alpha_t^{*, \lambda} = \tilde \al _t^{\lambda, \gamma^*(\lambda)}$ and the optimal control for $V_0(\lambda)$ is  thus 
    \beqs
    	\alpha_t^{*, \lambda} = \tilde{a}^{\lambda, \gamma^*(\lambda)}(t,X_t^{\tilde{a}^{\lambda, \gamma^*(\lambda)}},\hat B_t) = a_0^{Bayes,\lambda}(t,X_t^{{\al}^{*, \lambda}},\hat B_t), \;\;\; 0 \leq t \leq T,
    \enqs
 with the function $a_0^{Bayes,\lambda}$ as in $\eqref{eq: a_bayes}$.  
 This leads to the expression in \reff{abayes} from \reff{eq: Markowitz controls}, \reff{eq: atild} and \reff{eq: gammalambda}. Finally, the Lagrange multiplier 
 $\lambda(\vartheta)$ $=$ ${\rm arg}\min_{\lambda>0}\big[\lambda \vartheta - V_{0}(\lambda)\big]$ is explicitly computed from \reff{eq: V0lambda}, and is equal to the expression in  \reff{eq: explilambda}. The optimal performance of the Markowitz problem is then equal to
     \beqs
    	U_0(\vartheta) &=&  \lambda(\vartheta)\vartheta - V_0\big(\lambda(\vartheta) \big)  \; = \;  x_0 + \sqrt{ \vartheta \left( e^{R(0,b_0)}-1\right)},
    \enqs
    by \reff{eq: explilambda} and \reff{eq: V0lambda}. The optimal Bayesian-Markowitz strategy is then given by $\CM = \alpha^{*, \lambda(\vartheta)}$ accor\-ding to Lemma \ref{lem: dual_transform}. 
  \ep

\begin{remark}[Financial interpretation]
{\rm When the drift $B$ $=$ $b_0$ is known, the function $R$ is simply given by $R(t)$ $=$ $|\sigma^{-1}b_0|^2(T-t)$, and we retrieve the results recalled in Remark \ref{rem: constant_parameter}. Under a prior probability distribution on the drift $B$, we see that the optimal performance of the Bayesian-Markowitz pro\-blem has a similar form as in the constant drift case, when substituting $|\sigma^{-1}b_0|^2T$ by $R(0,b_0)$. For the optimal Bayesian-Markowitz strategy, we have to substitute the  vector term $\Sigma^{-1}b_0$ by $\Sigma^{-1}\hat B_t$, i.e. replacing $b_0$ by the posterior predictive mean $\hat B_t$, and correcting with the additional term $(\psi \sigma^{-1})^\trp \gradb R(t,\hat B_t)$.  We call $R$ the Bayesian risk premium function.
}
\epR
\end{remark}

Our main result in Theorem \ref{th: theomain} is stated under the condition that $R$ exists sufficiently smooth. In the next paragraph, we give some sufficient conditions ensuring this regularity, thus the existence of $R$, and provide some examples.


 \subsection{On existence and smoothness of the Bayesian risk premium}

In this section we provide sufficient conditions ensuring the existence of the Bayesian risk premium function $R$, solution to the semi-linear PDE  \eqref{eq: PDE_w}. Note that the standard assumptions of existence and uniqueness that we can find in \cite{Ladyzhenskaia1968} do not apply here since the function $\tilde{F}$ defined in \eqref{eq: DefFtild} is not globally Lipschitz in $p$. The difficulty of our framework comes from the unboundedness of the domain and the quadratic growth in $p$ of the function $\tilde{F}$ which entails that the solution may grow quadratically.

  \begin{theorem} \label{th: Existence Solution}
     Suppose the following conditions hold true:
    \begin{itemize}
    \item $\forall (t,b)\in \mathcal{R}$, $\psi(t,b)$ and $\psi^{-1}(t,b)$ are bounded,
    \item The matrix $\gradb \psi$ exists and is bounded,

    \end{itemize}
    then there exists a classical solution $R$ $\in$ ${\cal C}^{1,2}({\cal R})$ 
    to PDE \eqref{eq: PDE_w}. In addition, $R(t,b)$ is at most quadrically growing in $b$ and $\gradb R(t,b)$ is at most linearly growing in $b$.
  \end{theorem}

\noindent {\bf  Proof.}
    We provide a sketch of the proof and details in \ref{proof: th: Existence Solution}. 
    We follow a similar approach as \cite*{pham2002smooth}  and \cite*{Benth2005}.Without loss of generality we consider the function $\tilde{R}:=-R$. 
We rewrite the PDE characterizing $\Rtil$ as
  \beqs \label{PDE}
    	- \dt{\Rtil} - \frac{1}{2}\text{tr}(\psi\psi^\trp\Hessb\Rtil) + F(t,b,\gradb \Rtil) \;=\;  0
    \enqs
with terminal condition $\Rtil(T,.)$ $=$ $0$, where  the function $F$ $:$ ${\cal R}\times \Rn \to \R$ defined by
    \beqs
    	F(t,b,p) \;= \;  \frac{1}{2}|\psi^\trp p|^2 + 2 \big( \psi \sigma^{-1} b \big)^\trp p + |\sigma^{-1}b|^2,
    \enqs
 is quadratic (hence convex in the gradient argument $p$). Let us introduce the Fenchel-Legendre transform of $F(t,b,.)$ as 
    \beqs
    	L_t(b,q) \; := \; \max_{p \in \Rn} \big[  -q^\trp \psi^\trp p-F(t,b,p) \big], \;\;\; b \in {\cal B}_t, \; q \in \Rn,  
    \enqs
 	which explicitly yields   
   \beq \label{expliL}
    	L_t(b,q) \; = \;  \frac{1}{2}|q|^2 + 2(\sigma^{-1}b)^\trp q + |\sigma^{-1} b|^2.
    \enq
    The explicit form shows that $L_t$ only depends on $t$ through the domain ${\cal B}_t$ of $b$. 
It is also known that the following duality relationship holds:
    \beqs
    	F(t,b,p) \;=\; \max_{q \in \Rn} \big[ -q^\trp \psi^\trp p-L_t(b,q) \big].
    \enqs

From \reff{expliL} we deduce the following estimates on $L$ and its gradient w.r.t. $b$:
 \beq \label{estimL}
 \left\{
 \begin{array}{rcl}
 |L_t(b,q)| &\leq&  C (|q|^2  + |b|^2), \\
 |\nabla_b L_t(b,q)| &\leq& C (|q| + |b|), \;\;\; \forall b \in {\cal B}_t, \; q \in \R^n,
 \end{array}
 \right.
 \enq
 for some independent of $t$ positive constant $C$. Let us now consider the truncated auxiliary function
      \beqs
          F^k(t,b,p) \;=\; \max_{|q| \leq k} \big[ -q^\trp \psi^\trp p-L_t(b,q) \big], \;\;\; b \in {\cal B}_t, \; p \in \R^n, 
      \enqs
for $k$ $\in$ $\N$. By \reff{estimL} and Theorem 4.3 p. 163 in \cite*{Fleming2006}, there exists a unique smooth solution $\Rtil^k$ with quadratic growth condition to the truncated 
semi-linear PDE
   \beqs
 - \dt{\Rtil^k} - \frac{1}{2}\text{tr}( \psi\psi^\trp \Hessb \Rtil^k) + F^k(t,b,\gradb \Rtil^k) \;=\;  0,  
    \enqs
 with terminal condition $\Rtil^k(T,.)$ $=$ $0$.
 
The next step is to obtain estimates  on $\tilde{R}^k$ and its gradient w.r.t. $b$, uniformly in $k$:
 \beqs \label{estimR}
 \left\{
 \begin{array}{rcl}
 |\Rtil^k(t,b)| &\leq&  C (1  + |b|^2), \\
 |\nabla_b \Rtil^k(t,b)| &\leq& C (1 + |b|), \;\;\; \forall (t,b) \in {\cal R},
\end{array}
 \right.
 \enqs
 for some constant $C$ independent of $k$. Then, following similar arguments as in \cite*{pham2002smooth} and \cite*{Benth2005}, we show that for $k$ large enough, 
 $R^k =-\Rtil^k$ solves PDE \eqref{eq: PDE_w}. This proves the existence of a smooth solution to this semi-linear PDE. 
 \ep

  \subsection{Examples}

We illustrate with some relevant examples  the explicit computation of the diffusion coefficient  $\psi$ appearing in the dynamic of the posterior mean of the drift described in \eqref{eq: dhatB psi},  as well as the computation of  the Bayesian risk premium function $R$.


  \subsubsection{Prior discrete law} 
  
 We consider the case when the drift vector $B$ has a prior discrete distribution
 \beq \label{eq: multi-point law}
  \mu(db) &=&  \sum_{i=1}^{N} \pi_i\delta_{V_{i}}(db),
 \enq 
 where for $i$ $=$ $1,\ldots,N$, $V_i$ are vectors in $\R^n$, $\pi_i$ $\in$ $(0,1)$ and $\sum_{i=1}^N \pi_i$ $=$ $1$.   
 We denote by $V$ the $n\times N$-matrix $V$ $=$ $(V^1 \ldots V^N)$, with rank(V) = $n < N$ and we assume that the vectors $V_{i}$ are chosen such that 
  \beqs 
  	{\rm Cov}(B)  = \sum_{i = 1}^N \pi_i V_iV_i^\trp-b_0b_0^\trp>0.
  \enqs
\vspace{2mm}
 
 From   \reff{eq: multi-point law} and \reff{eq: mu(t,y)}, we easily compute the conditional distribution  of $B$  w.r.t. $Y_t$ $=$ $y$ $\in$ 
 $\R^n$, which is given by
  \beqs \label{eq: mu multi-point}
    	\mu_{t,y}(db) \;= \;  \sum_{i=1}^{N}p^i_{t,y} \delta_{V_i}(db) \quad t \in [0,T],
    \enqs
    where $p_{t,y}$ $=$ $(p_{t,y}^1 \ldots p_{t,y}^N)$ $\in$ $[0,1]^N$ is determined by 
    \beqs
 p^i_{t,y} & = & 
  \frac{\pi_i e^{y^\trp\sigma^{-1}V_i-\frac{1}{2}V_i^\trp\Sigma^{-1}V_it}}{\sum_{j=1}^{N}\pi_j e^{y^\trp\sigma^{-1}V_j-\frac{1}{2}V_j^\trp\Sigma^{-1}V_jt}}, 
  \quad  i =1,\ldots,N.  
  \enqs
It follows that the function $f$ in \reff{eq: f(t,y)} is equal to
 \beqs
 f(t,y) \; = \;  \int_{\Rn}b \mu_{t,y}(db)  &=&  \sum_{i=1}^{N} p^i_{t,y} V_i \; = \; V p_{t,y}, 
 \enqs
from which we calculate its gradient:
\beqs
\nabla_y f(t,y) &=&  \left(\sum_{i=1}^{N}p^i_{t,y} V_i V_i^\trp - V p_{t,y} (V p_{t,y})^\trp \right) (\sigma^{-1})^\trp. 
\enqs   
In this case the domain $\mathcal{B}_t$ of $b$ is the convex hull of the vectors $V_i$, $i=1, \dots, N$ mathematically defined as
\beqs 
	\mathcal{B}_t = \left\{ b : Vp_{t,y^*}=b \text{ and } \forall j \in \{1, \ldots ,N\}, \;\; p^j_{t,y^*} \in (0,1) \; \text{ and } \sum_{j=1}^Np^j_{t,y^*}=1 \right\}, 
\enqs 
where for fixed $(t,b)$ $\in$ ${\cal R}$,  $y^*$ $=$ $f_t^{-1}(b)$. More precisely, $y^*$ is the solution to the equation 
$f(t,y^*)$ $=$ $V p_{t,y^*}$ $=$ $b$. We note $\overline{\mathcal{B}}_t$ the closure of $\mathcal{B}_t$.
\\
The function $\psi$ defined in \reff{eq: psi} is therefore equal to
\beq
\begin{aligned} \label{eq: explipsidiscret}  
\psi(t,b) &=  \nabla_y f(t,y^*) \\
       & =  \left(\sum_{i=1}^{N}p^i_{t,y^*} V_i V_i^\trp - b b^\trp \right) (\sigma^{-1})^\trp. \\    
\end{aligned}
\enq
   
\begin{remark}   
{\rm In the discrete case we cannot apply Theorem \ref{th: Existence Solution} for two main reasons: $\mathcal{B}_t$ is a bounded non-smooth domain, and $\psi$ can vanish on the boundary of $\mathcal{B}_t$, leading to $\psi^{-1}$ being ill defined on $\overline{\mathcal{B}_t}$. It is, in particular, easy to see that $\psi$ vanishes at the points $V_i$. Indeed, $b$ equals $V_i$ implies that $\forall$ $j =1, \dots, N$, $p^j_{t,y^*}=0$ for $j \neq i$ and $p^i_{t,y^*}=1$ leading to Eq. \eqref{eq: explipsidiscret} being zero. The existence theorem of a solution to problem \reff{eq: PDE_w}-\reff{eq: termcond}, in our context, can be found in \cite*{Krylov1987}, especially chapter 6, section 3.}
\epR
\end{remark}


  \subsubsection{The Gaussian case} \label{subsec: Gaussian_Case}
  
  We consider the case when the drift vector $B$ follows a $n$-dimensional normal distribution with mean vector $b_0$ and covariance matrix $\Sigma_0$:
  \beq \label{eq: gaussianmu}
  \mu & \sim & {\cal N}(b_0,\Sigma_0).
  \enq
 In this conjugate case, it is well-known (see Proposition 10 in \cite*{Gueant2017}) that the posterior distribution of $B$, i.e., the conditional distribution of $B$ given $Y_t$ $=$ $y$ is also Gaussian with mean 
 \beq  \label{fmean}
 f(t,y) & = &   \Big(\Sigma_0^{-1}+\Sigma^{-1}t \Big)^{-1} \Big( \Sigma_0^{-1} \mB+(\sigma^{\trp})^{-1}y \Big),
 \enq
 and covariance
 \beqs
 \Sigma(t,y) &=& \Big( \Sigma_0^{-1}+\Sigma^{-1}t \Big)^{-1}.
 \enqs
 For the sake of completeness, we show this result. From  \reff{eq: mu(t,y)} and \reff{eq: gaussianmu},  
 the conditional distribution  of $B$  w.r.t. $Y_t$ $=$ $y$ is given by 
  \beqs
         \mu_{t,y}(db) &= & \frac{e^{ y^\trp\sigma^{-1}b - \frac{1}{2}b^\trp\Sigma^{-1}bt - \frac{1}{2}(b-\mB)^\trp\Sigma_0^{-1}(b-\mB)}}{\int_{\Rn} e^{ y^\trp \sigma^{-1}b - \frac{1}{2}b^\trp\Sigma^{-1}bt - \frac{1}{2}(b-\mB)^\trp\Sigma_0^{-1}(b-\mB)}db}db \\
        &= & \frac{e^{ - \frac{1}{2} \left( b^\trp\left( \Sigma_0^{-1}+\Sigma^{-1}t  \right)b -2b^\trp\left( \Sigma_0^{-1} \mB + (\sigma^{-1})^\trp y\right)\right)}}{\int_{\Rn} e^{ - \frac{1}{2} \left( b^\trp\left( \Sigma_0^{-1}+\Sigma^{-1}t \right)b -2b^\trp\left(\Sigma_0^{-1} \mB + (\sigma^{-1})^\trp y\right)\right)}db}db \\
       & = & \frac{e^{ - \frac{1}{2} \left( b-\left( \Sigma_0^{-1}+\Sigma^{-1}t  \right)^{-1}(\Sigma_0^{-1} \mB + (\sigma^{-1})^\trp y)\right)^\trp \left( \Sigma_0^{-1}+\Sigma^{-1}t  \right)\left( b-\left( \Sigma_0^{-1}+\Sigma^{-1}t \right)^{-1}(\Sigma_0^{-1} \mB + (\sigma^{-1})^\trp y)\right)}}{\int_{\Rn} e^{ - \frac{1}{2} \left( b-\left( \Sigma_0^{-1}+\Sigma^{-1}t  \right)^{-1}(\Sigma_0^{-1} \mB + (\sigma^{-1})^\trp y)\right)^\trp \left( \Sigma_0^{-1}+\Sigma^{-1}t \right)\left( b-\left( \Sigma_0^{-1}+\Sigma^{-1}t  \right)^{-1}(\Sigma_0^{-1} \mB + (\sigma^{-1})^\trp y)\right)}db}db. \\
\enqs
Recognising the form of a Gaussian distribution, we find
    \beqs
         \mu_{t,y}(db) &=& \frac{e^{ - \frac{1}{2} \left( b-\left( \Sigma_0^{-1}+\Sigma^{-1}t  \right)^{-1}(\Sigma_0^{-1} \mB + (\sigma^{-1})^\trp y)\right)^\trp \left( \Sigma_0^{-1}+\Sigma^{-1}t  \right)\left( b-\left( \Sigma_0^{-1}+\Sigma^{-1}t  \right)^{-1}(\Sigma_0^{-1} \mB + (\sigma^{-1})^\trp y)\right)}}{(2\pi)^{\frac{n}{2}} |\left( \Sigma_0^{-1}+\Sigma^{-1}t \right)^{-1}|^{\frac{1}{2}}}db,
 \enqs
 which shows that 
 \beqs
 \mu_{t,y} & \sim &   \mathcal{N} \left(f(t,y) ,\Sigma(t,y) \right). 
 \enqs  
 
Next,  from \reff{fmean}, we have
 \beqs
   \nabla_y f(t,y) &=&  \big( \Sigma_0^{-1}+\Sigma^{-1}t  \big)^{-1}(\sigma^{-1})^\trp,
 \enqs
Left multiplying by $\Sigma_0 \Sigma_0^{-1}$ and right multiplying by $\sigma^\trp \Sigma^{-1} \sigma$  the previous equation, we obtain the multi-dimensional independent from $b$ familiar expression of the diffusion coefficient $\psi$ stated in \cite*{Ekstrom2016} in dimension one:
 \beqs
  \psi(t) &=& \Sigma_0\big( \Sigma+\Sigma_0 t \big)^{-1} \sigma.
 \enqs

 For the computation of the Bayesian risk premium function $R$ solution to   \reff{eq: PDE_w}, we look for a solution of the form: 
  \beq \label{Rquadra}
 R(t,b) &=& b^\trp M(t)b + U(t),
\enq
for some  $\R^{n\times n}$-valued  function $M$, and $\R^n$-valued function $U$ defined on $[0,T]$, to be determined. Plugging into \reff{eq: PDE_w}, we see that $M$ and $U$ should satisfy the first order ODE system: 
    \beq \label{eq: M Gaussian}
      \left\{
      \begin{array}{rcl}
        -M'(t) - 2M(t)^\trp G(t)^{\trp} \Sigma G(t)M(t) + 4 G(t) M(t) - \Sigma^{-1} &=& 0 \\
        -U'(t) - \text{tr} \big[ G(t)^{\trp} \Sigma G(t) M(t) \big]  &=& 0,
      \end{array}
      \right.
    \enq
with $G(t)$ $=$ $(\Sigma+\Sigma_0 t)^{-1}\Sigma_0$  and terminal conditions:    
    \beq \label{eq: M Gaussian term cond}
        M(T) \; = \;  0, & & 
        U(T) \; =\;  0.
    \enq

The solution to \eqref{eq: M Gaussian}-\eqref{eq: M Gaussian term cond} is given by the following Lemma, whose proof is detailed in \ref{proof lem: sol Gaussian Multi}.
 
   \begin{lemma} \label{lem: sol Gaussian Multi}
      The solution to the ODE system \reff{eq: M Gaussian} with terminal condition \reff{eq: M Gaussian term cond} is:
      \beqs
      M(t) & = &  \Sigma_0^{-1}+\Sigma^{-1}t-\left[  \left( \Sigma_0^{-1}+\Sigma^{-1}T\right)^{-1}+2 \int_t^T \Sigma_0 \left( \Sigma+ \Sigma_0 s\right)^{-1} \Sigma \left( \Sigma + \Sigma_0 s\right)^{-1} \Sigma_0 ds\right ]^{-1} \\
      U(t) &=&   \int_t^T {\rm tr}\left( \Sigma_0 \left( \Sigma + \Sigma_0 s\right)^{-1} - \left[  \left( ( \Sigma_0^{-1}+ \Sigma^{-1}T) ^{-1} +2 \int_s^T G(u)^\trp \Sigma G(u)du\right) \left( G(s)^\trp \Sigma G(s) \right)^{-1} \right]^{-1}  \right)ds
      \enqs
    \end{lemma}

\vspace{5mm}

We conclude that the Bayesian risk premium function $R$ is explicitly given by \reff{Rquadra} with $M$ and $U$ as in Lemma  \ref{lem: sol Gaussian Multi}. 
 The optimal Bayesian mean-variance strategy \reff{abayes} is explicitly written as 
 \beqs
 \alpha_t^{*,\lambda}  &=&  \Big(x_0- X_t^{ \alpha^{*,\lambda}} +\frac{e^{R(0,b_0)}}{2\lambda}\Big) 
 \left[ \Sigma^{-1} - \Big(  \Sigma + \Sigma_0 t \big)^{-1} \Sigma_0  M(t) \right] \hat B_t
    \enqs   
 where $\hat B_t$ $=$ $f(t,Y_t)$ $=$ $f(t,h(t,S_t))$.

\vspace{2mm}

In the one-dimensional case $n$ $=$ $1$, hence with $\Sigma$ $=$ $\sigma^2$, $\Sigma_0$ $=$ $\sigma_0^2$, the expressions for $M$ and $U$ are simplified into
 \beqs \label{eq: M1}
  M(t) &=&  \frac{\sigma^2+\sigma_0^2 t}{\sigma^2 \big( \sigma^2+\sigma_0^2 (2T-t)\big)}(T-t), \\
   U(t) &=&  \ln \left( \frac{\sigma^2+\sigma_0^2T}{\sqrt{(\sigma^2+\sigma_0^2t)(\sigma^2+\sigma_0^2(2T-t))}}\right).
\enqs  


\section{Impact of learning on the Markowitz strategy} \label{sec: Impact_Learning}

This section shows the benefit of using a Bayesian learning approach to solve the Markowitz problem. In contrast to classical approaches that estimate unknown parameters in a second step, leading to sub-optimal solutions, the Bayesian learning approach uses the updated data from the most recent prices of the assets to adjust the controls of the investment strategy and reaches optimality in the Markowitz sense. 
\\

In a framework where the drift $B$ is unknown, following a prior probability  distribution $\mu$,  we will compare the performance, in terms of Sharpe ratio, of the Bayesian learning strategy to the non-learning strategy. Recall, the non-learning strategy considers the drift constant set at $b_0$ $=$ $\E[B]$ $=$ $\int b \mu(db)$. This allows us to exhibit the benefit of integrating the most recent available information into the strategy.


\subsection{Computation of the Sharpe ratios}

The Sharpe ratio associated to  a portfolio strategy $\alpha$ is defined by
\beqs
Sh_T^\alpha & := & \frac{\E[X_T^\alpha] -x_0}{\sqrt{{\rm Var}(X_T^\alpha)}}. 
\enqs
We denote by $\hat \alpha^{\vartheta,L}$ the optimal Bayesian-Markowitz strategy obtained in Theorem \ref{th: theomain}, $X^L = X^{\hat \alpha^{\vartheta,L}}$ the associated wealth process and $Sh_T^{L}$ the corresponding Sharpe ratio.  
We also write $\hat \alpha^{\vartheta, NL}$ the non-learning Markowitz strategy based on a constant drift parameter $b_0$ as described in  Remark \ref{rem: constant_parameter}, $X^{NL}=X^{\hat \alpha^{\vartheta,NL}}$ the associated wealth process and 
$Sh_T^{NL}$ the corresponding Sharpe ratio. 

\begin{prop} \label{propsharpe}
                   
The Sharpe ratios of the learning and non-learning Markowitz strategies are explicitly given by
\beq
Sh_T^{L} &=& \sqrt{e^{R(0,b_0)}-1}, \label{ShL}\\
Sh_T^{NL} &=& 
\frac{1-\int_{\Rn}e^{-b_0^\trp\Sigma^{-1}bT}\mu(db)}{\sqrt{ \int_{\Rn}e^{ -b_0^\trp\Sigma^{-1}(2b-b_0)T} \mu(db)-\left( \int_{\Rn}e^{-b_0^\trp\Sigma^{-1}bT} \mu(db)\right)^2}}. \label{ShNL}
\enq
\end{prop}
\noindent {\bf Proof.}

{\bf 1.} We first focus on the Bayesian learning strategy. From \eqref{eq: optterminal}, we have  $\E[X^L]$ $=$ 
$x_0+\frac{1}{2 \lambda(\vartheta)}\left( e^{R\left(0,b_0 \right)}-1\right)$ with $\lambda(\vartheta) = \sqrt{\frac{e^{R \left(0,b_0 \right)}-1}{4\vartheta}}$ binding the variance of the optimal terminal wealth to $\vartheta$.   We thus  obtain 
    \beqs
    Sh^{L}_T &=& 
	\frac{\E[X^L]-x_0}{\sqrt{\vartheta}} \;=\; \frac{\sqrt{\vartheta\left( e^{R\left(0,b_0 \right)}-1\right)}}{\sqrt{\vartheta}} \;=\;  \sqrt{ e^{R\left(0,b_0 \right)}-1}.
    \enqs
 
 \noindent {\bf 2.} Let us now consider the non-learning strategy $\hat \alpha^{\vartheta,NL}$ for which we need to compute its expectation and variance.  
 From the expression of $\hat \alpha^{\vartheta,NL}$ given  in 
 Remark \ref{rem: constant_parameter}, and the self-financed equation of the wealth process \reff{eq: process_X},  
 we deduce the dynamic of the conditional expectation of $ X^{NL}$ given $B$:
 \beqs
  d\E[  X_t^{NL}|B] &= & \E\Big[ \big( x_0-  X_t^{NL} +C_1 \big)C_2^\trp B|B \Big]dt \\
        & = & \Big[ \left( x_0+C_1 \right) C_2^\trp B -\E\big[  X_t^{NL} |B \big] C_2^\trp B \Big] dt,
 \enqs
where we set $C_1 = \frac{e^{|\sigma^{-1}b_0|^2T}}{2 \lambda_0(\vartheta)}$, $C_2 =  \Sigma^{-1}b_0$ to alleviate notations. 
We thus obtain  an ordinary diffe\-rential equation for  $t \mapsto \E[X_t^{NL} | B]$ with initial condition $\E[ X_0^{NL} |B]$ $=$ $x_0$, whose solution is explicitly given by 
    \beq \label{XNL}
   \E[ X^{NL}_t |B] &=&  x_0 + C_1 \big(1-e^{-C_2^\trp Bt}\big), \;\;\; 0 \leq t \leq T. 
    \enq
Integrating w.r.t. the law of $B$, we obtain the expression of the expectation:  
\beq \label{EXNL}
\E[ X_t^{NL} ] &=& x_0 + \frac{e^{||\sigma^{-1}b_0||^2T}}{2 \lambda_0(\vartheta)} \left( 1-\int_{\Rn}e^{-b_0^\trp\Sigma^{-1}bt}\mu(db)\right). 
\enq
 
Next, let us compute the variance of $ X_T^{NL}$. Applying It\^o's formula to $| X^{NL}|^2$, and taking conditional expectation w.r.t. $B$,  we have 
    \beqs
    	d\E[ | X_t^{NL}|^2|B] &=&  \Big[ 
	C_2^\trp \left(\Sigma C_2-2B \right)\E[|X_t^{NL}|^2|B]-2C_2^\trp \left(x_0+C_1 \right) \left(\Sigma C_2 -B\right)\E[X_t^{NL} |B]  \\
	& & \;\;\;\;\; + 	\;C_2^\trp \Sigma C_2 \left( x_0 + C_1 \right)^2 \Big] dt.  
    \enqs
    Replacing $\E[ X^{NL}_t |B]$ by its value calculated in \reff{XNL}, we obtain an ODE in $\E[|X_t^{NL}|^2 |B]$ that we explicitly solve: 
    \beqs
    	\E[| X_t^{NL}|^2 |B] &=& \left(x_0+C_1 \right)^2 -2C_1 \left( x_0+C_1\right)e^{C_2^\trp Bt}+C_1^2e^{C_2^\trp \left( \Sigma C_2 -2B\right)t}.
    \enqs
By integrating over $B$, we obtain
    \beqs
    	\E[|X_t^{NL}|^2 ] &=&  \left(x_0+C_1 \right)^2 -2C_1 \left( x_0+C_1\right) \int_{\Rn}e^{C_2^\trp bt} \mu(db)+C_1^2 e^{C_2^\trp \Sigma C_2t} \int_{\Rn}e^{-2C_2^\trp bt} \mu(db).
    \enqs
From ${\rm Var}(X_t^{NL})$ $=$ $\E[|X_t^{NL}|^2 ]$ $-$ $\big(\E[X_t^{NL}]\big)^2$, we obtain the expression of the variance
\beq \label{VXNL}
{\rm Var}(X_t^{NL}) &=& \frac{e^{2|\sigma^{-1}b_0|^2T}}{4 \lambda_0(\vartheta)^2} \left[ \int_{\Rn}e^{ -b_0^\trp\Sigma^{-1}(2b-b_0)t} \mu(db)-\left( \int_{\Rn}e^{-b_0^\trp\Sigma^{-1}bt} \mu(db)\right)^2\right].
\enq
Finally, coupling \reff{EXNL} and \reff{VXNL} we infer the value of $Sh_T^{NL}$ in \reff{ShNL}. 
\ep
 
\vspace{5mm}

Next we prove that $Sh_T^{NL}$ $\leq$ $Sh_T^L$. From \reff{eq: process_X} we know that $X^{NL}$ is linear in its control $\hat \alpha^{\vartheta, NL}$, so we can always find a leveraged strategy $\tilde \alpha^{\vartheta, NL} = \delta \hat \alpha^{\vartheta, NL}$ with $\delta >0$ such that ${\rm Var}(X_T^{\hat \alpha^{\vartheta, NL}}) = \vartheta$, simply by taking $\delta = \sqrt{\frac{\vartheta}{{\rm Var}({X}_T^{NL})}}$. Thus, by invariance of the Sharpe ratio w.r.t. the leverage we obtain,
\beqs
\begin{aligned}
	Sh^L_T-Sh^{NL}_T &= \frac{\E[{X}_T^{\hat \alpha^{\vartheta, L}}] -x_0}{\sqrt{\vartheta}}-\frac{\E[X_T^{\hat \alpha^{\vartheta, NL}}] -x_0}{\sqrt{{\rm Var}(X_T^{\hat \alpha^{\vartheta, NL}})}}\\
    &= \frac{\E[{X}_T^{\hat \alpha^{\vartheta, L}}] -x_0}{\sqrt{\vartheta}}-\frac{\E[X_T^{\tilde \alpha^{\vartheta, NL}}] -x_0}{\sqrt{\vartheta}}\\
    &= \E[{X}_T^{\hat \alpha^{\vartheta, L}}] - \E[X_T^{\tilde \alpha^{\vartheta, NL}}] \\
    & \geq 0
    \end{aligned}
\enqs
The last inequality comes from $\hat \alpha^{\vartheta, L}$ being the optimal control.
\begin{remark}
{\rm We have an upper-bound for the Sharpe ratio of the non-learning strategy:
\beqs \label{eq: upperNL}
Sh^{NL}_T  &\leq&  \sqrt{e^{|\sigma^{-1}b_0|^2T}-1}. 
\enqs
Indeed, by  Jensen's inequality, we have
\beqs
         \left( \int_{\Rn} e^{-b_0 \Sigma^{-1}bT} \mu(db) \right)^2 &\leq & \int_{\Rn} e^{-2b_0 \Sigma^{-1}bT} \mu(db),\\
        \int_{\Rn} e^{-b_0 \Sigma^{-1}bT} \mu(db) &\geq & e^{-T\int_{\Rn} b_0 \Sigma^{-1}b \mu(db) } = e^{-|\sigma^{-1} b_0 |^2T}, \\
        \int_{\Rn} e^{-2b_0 \Sigma^{-1}bT} \mu(db) &\geq & e^{-2|\sigma^{-1} b_0 |^2T}. \\
\enqs
From \reff{ShNL}, we thus deduce
     \beqs
         Sh^{NL}_T &\leq& \frac{1-\int_{\Rn} e^{-b_0 \Sigma^{-1}bT} \mu(db)}{\sqrt{\int_{\Rn} e^{-b_0 \Sigma^{-1}(2b-b_0)T} \mu(db)-\int_{\Rn} e^{-2b_0 \Sigma^{-1}bT} \mu(db)}} \\
        & = &\frac{1-\int_{\Rn} e^{-b_0 \Sigma^{-1}bT} \mu(db)}{\sqrt{\left( e^{|\sigma^{-1}b_0|^2T}-1 \right)\int_{\Rn} e^{-2b_0 \Sigma^{-1}bT} \mu(db)}}\\
        &\leq & \frac{1-e^{-|\sigma^{-1}b_0|^2T}}{\sqrt{\left( e^{|\sigma^{-1}b_0|^2T}-1 \right)e^{-2|\sigma^{-1}b_0|^2T}}} \; = \; \sqrt{e^{|\sigma^{-1}b_0|^2T}-1 }
    \enqs

We notice that when $\mu = \mathbbm{1}_{b_0}$ the Sharpe ratio of the non-learning strategy coincides with $\sqrt{e^{|\sigma^{-1}b_0|^2T}-1}$. 
}
\epR
\end{remark} 
 
\subsection{Information Value} \label{subsec: Info value}

We illustrate our results in the case where $B$ follows a prior one-dimensional Gaussian distribution ${\cal N}(b_0,\sigma_0^2)$.    
To assess the value of information (VI), we consider the Sharpe ratio $Sh_T^{L}$ and $Sh_T^{NL}$ of the learning and non-learning strategies as computed in Proposition \ref{propsharpe}. We define the value of information as the difference $Sh_T^L- Sh_T^{NL}$, and measure its sensitivity w.r.t. various parameters. So, in the sequel, we denote VI(Asset i) the value of information obtained using the set of parameters corresponding to Asset i. 
\\

The following explicit formulas for the Sharpe ratios learning and non-learning, used to create the graphs of this section, are computed from \reff{ShL} and \reff{ShNL} with $\mu \sim \mathcal{N}(b_0, \sigma_0^2)$. 
\beq \label{eq: Gauss: ShL}
	Gaussian\_Sh^L_T = \sqrt{\frac{\sigma^2+\sigma_0^2T}{\sigma \sqrt{\sigma^2 + 2\sigma_0^2T}}e^{\frac{b_0^2T}{\sigma^2+2\sigma_0^2T}}-1}
\enq 
and
\beq \label{eq: Gauss: ShNL}
	Gaussian\_Sh^{NL}_T = \frac{e^{\frac{b_0^2}{\sigma^2}T\left(1-\frac{\sigma_0^2}{2\sigma^2}T \right)}-1}{\sqrt{e^{\frac{b_0^2}{\sigma^2}T \left(1+\frac{\sigma_0^2}{\sigma^2}T\right)}-1}}
\enq 

\subsubsection{Volatility of the drift}
Intuitively, the volatility of the drift $\sigma_0$ measures the confidence the investor puts in her estimate of $b_0$. The higher $\sigma_0$, the more confident the investor is about her estimate. 
It is an important parameter since it legitimates the use of a learning strategy. 
\\

\begin{figure}[H]
\centering
  \includegraphics[scale=0.25]{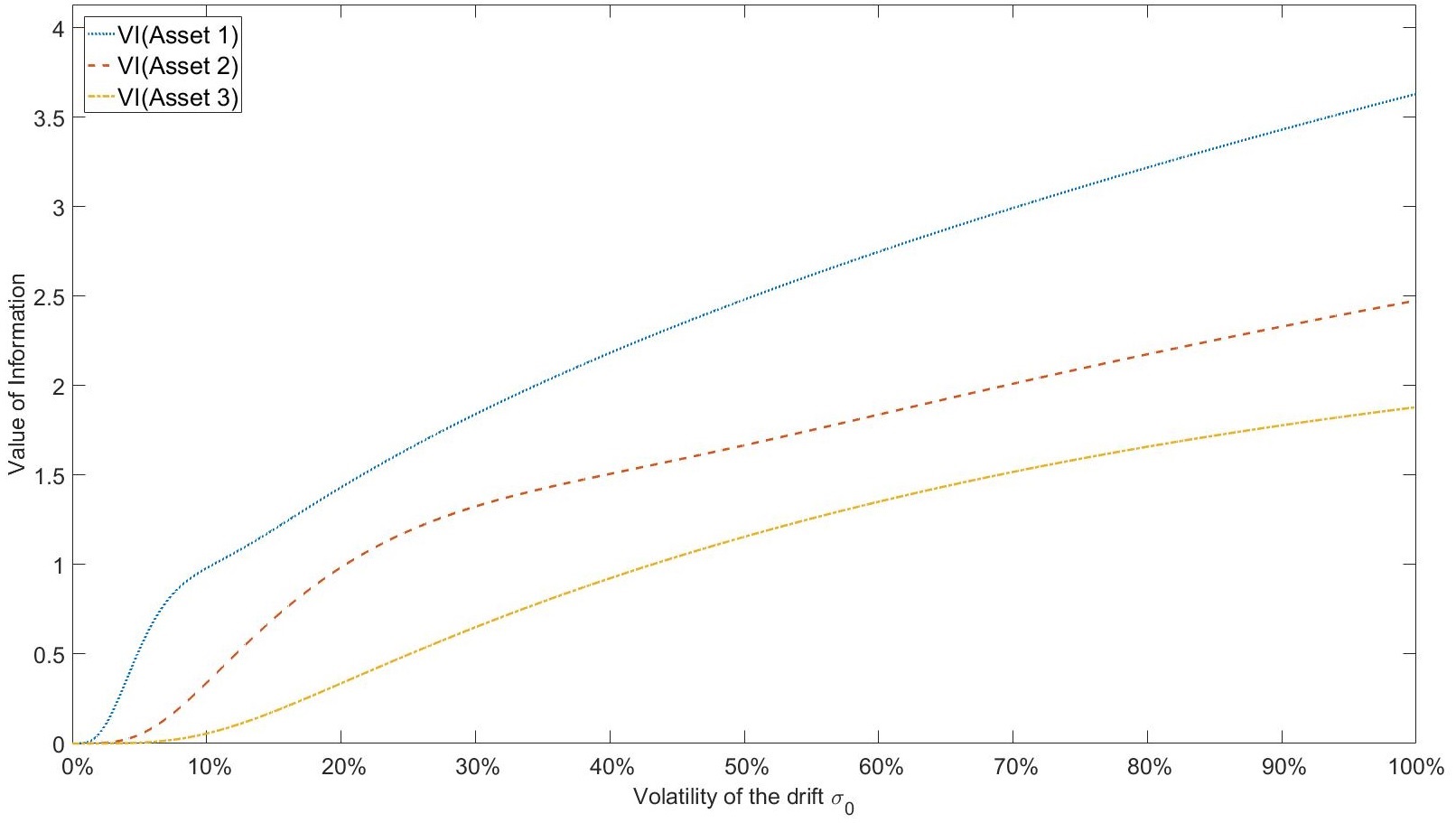}  
  \caption{Value of information as a function of $\sigma_0$ with parameters as in Table \ref{tab: F1 parameters}. }
  \label{graph: drift_volatility}
\end{figure}

To estimate the sensitivity of the value of information w.r.t. the volatility of the drift, we choose an investment horizon of $T=1$ and a sample of three different assets with the same mean $b_0 = 5\%$ and different volatilities, $\sigma =\{5\%, 10\%$, $20\%\}$ resulting in realistic Sharpe ratios of $1$, $0.5$ and $0.25$ respectively.
From Formulas \eqref{eq: Gauss: ShL} and \eqref{eq: Gauss: ShNL} we compute the value of information according to the volatility of the drift ranging from $0$ to $100\%$.
\\
Figure \ref{graph: drift_volatility} shows the value of information as a function of the volatility of the drift for the three assets described previously, which parameters are summed up in Table \ref{tab: F1 parameters}. 
As we can see on the graph, the value of information is a monotone increasing function of the volatility of the drift. When the volatility of the drift is zero, as in the Dirac case, the Sharpe ratios of the learning and non-learning strategies are equal. 
It is clear since a volatility of the drift equal to zero means the drift is simply constant, so updating it does not bring any additional value to the learning strategy. 
For all three assets, we notice that as $\sigma_0$ increases the value of information becomes more and more valuable depending on the level of the Sharpe ratio of each asset. 
The higher the Sharpe ratio, the bigger the value of information and its rate of increase. For instance, the value of information of the highest Sharpe ratio asset, Asset 1, increases rapidly to $1$ when $\sigma_0 = 10\%$ and reaches rapidly $3.5$ when $\sigma_0=100\%$ whereas the lowest Sharpe ratio asset, Asset 3, equals roughly $0.1$ when $\sigma_0 = 0.1$ and nearly reaches $2$ when $\sigma_0=100\%$ at a relatively slower pace.

\begin{table}[H]
\caption{Parameter values used in Figure \ref{graph: drift_volatility}.}
      \centering 
      \begin{tabular}{cccccc} 
        \cline{2-6}
          \multicolumn{1}{c}{}&$\mathbf{b_0}$ & $\boldsymbol{\sigma}$ & \textbf{Sharpe ratio} & \textbf{T} &  $\boldsymbol{\sigma}_\mathbf{0}$\\
         \hline
         $\mathbf{Asset\;1}$ &5\% & 5\% & 1 & 1 &  [0-100\%] \\
		
        $\mathbf{Asset\;2}$ &5\% & 10\% & 0.5 & 1 &  [0-100\%] \\
        
         $\mathbf{Asset\;3}$ &5\% & 20\% & 0.25 & 1 & [0-100\%] \\
        \hline
      \end{tabular}     
      \label{tab: F1 parameters}
\end{table}

\begin{figure}[H]
\centering
\includegraphics[scale=0.25]{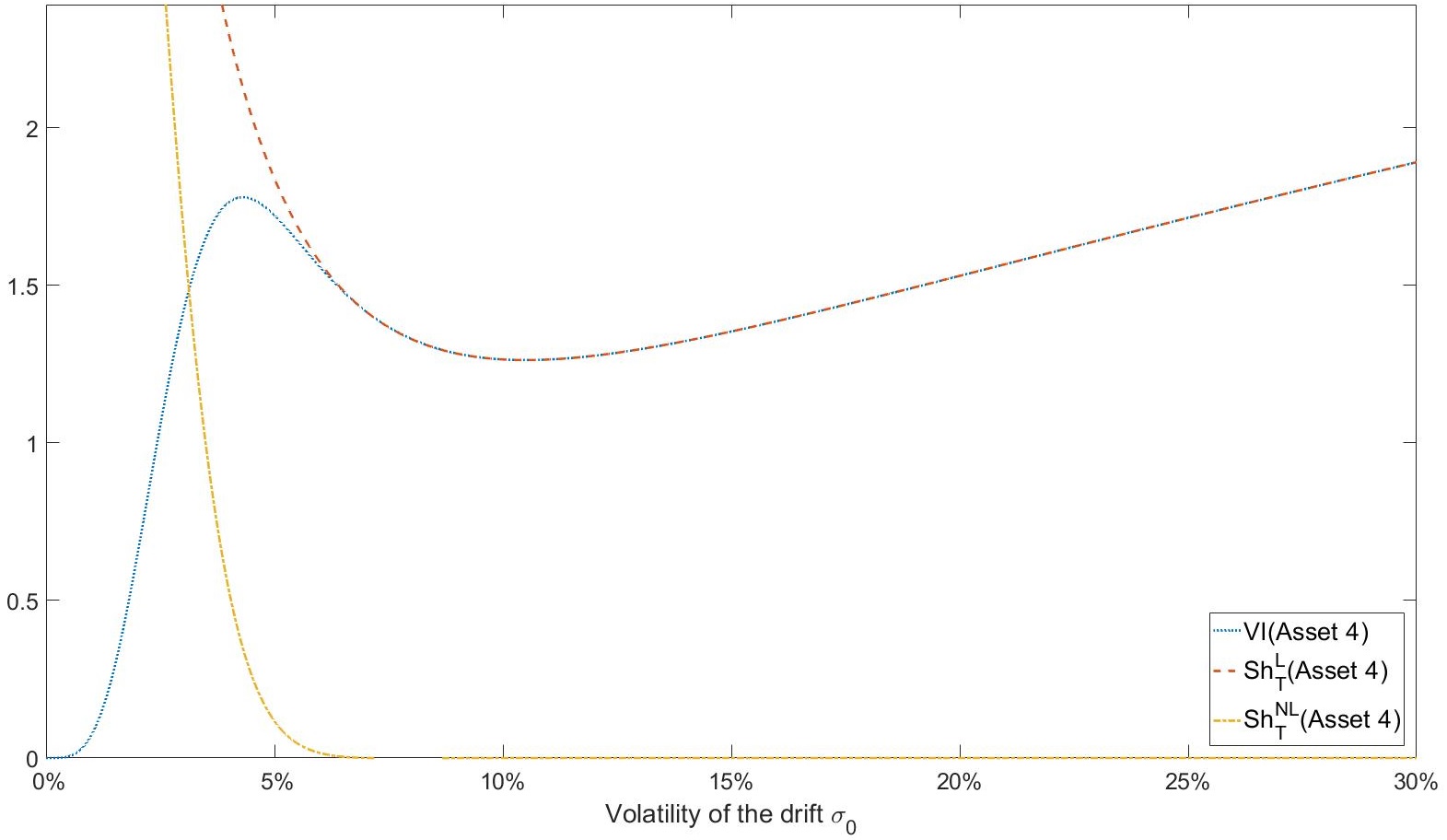}
	\caption{Value of information, $Sh^L_T$ and $Sh^{NL}_T$ of Asset 4 as a function of $\sigma_0$ with parameters as in Table \ref{tab: F2 parameters}.}
  \label{graph: drift_volatility_anomaly}
\end{figure}

An interesting fact happens in the case of an asset with a high Sharpe ratio. Figure \ref{graph: drift_volatility_anomaly} shows the value of information, the curves of the Sharpe ratio of the learning and the non-learning strategy based on Asset 4. This asset has a high Sharpe ratio of $2$, a set of parameters as in Table \ref{tab: F2 parameters} and a volatility of the drift ranging from $0$ to $30\%$.
\\

\begin{table}[H]
    \centering       
	\caption{Parameter values used in Figure \ref{graph: drift_volatility_anomaly}.}
	\begin{tabular}{cccccc} 
        \cline{2-6}
          \multicolumn{1}{c}{}&$\mathbf{b_0}$ & $\boldsymbol{\sigma}$ & \textbf{Sharpe ratio} & \textbf{T} &  $\boldsymbol{\sigma}_\mathbf{0}$\\
         \hline
         $\mathbf{Asset\;4}$ &10\% & 5\% & 2 & 1&  [0-30\%]\\
        \hline
      \end{tabular}
      \label{tab: F2 parameters}
\end{table}
As we can see on the graph, the value of information curve is no more monotonic and we observe a bump around $\sigma_0 = 5\%$. To explain this shape, we see that when $\sigma_0$ ranges between $0$ and $5\%$, both Sharpe ratios decrease. 
It is understandable since the variance of the terminal wealth is an increasing function of $\sigma$ and $\sigma_0$. 
Nevertheless, the decrease is much more rapid for the non-learning strategy reaching approximately zero when the drift volatility merely exceeds $5\%$, approximately the value of the volatility of the asset. 
The difference in the rate of decrease explains the bump we observe. Then, the Sharpe ratio of the learning strategy keeps decreasing and progressively recover when the volatility of the drift approaches the double of the volatility of the asset, $10\%$ . At this point, the learning effect really allows the strategy to exploit its advantage of updating the drift as the new information become available. As soon as $\sigma_0 >2\sigma$, the signal coming from the volatility of the drift differentiates from the volatility of the asset and is captured by the learning strategy which allows the value of information to increase.

\subsubsection{Sharpe ratio of the asset} 

\begin{figure}[H]
	\centering
  \includegraphics[scale=0.25]{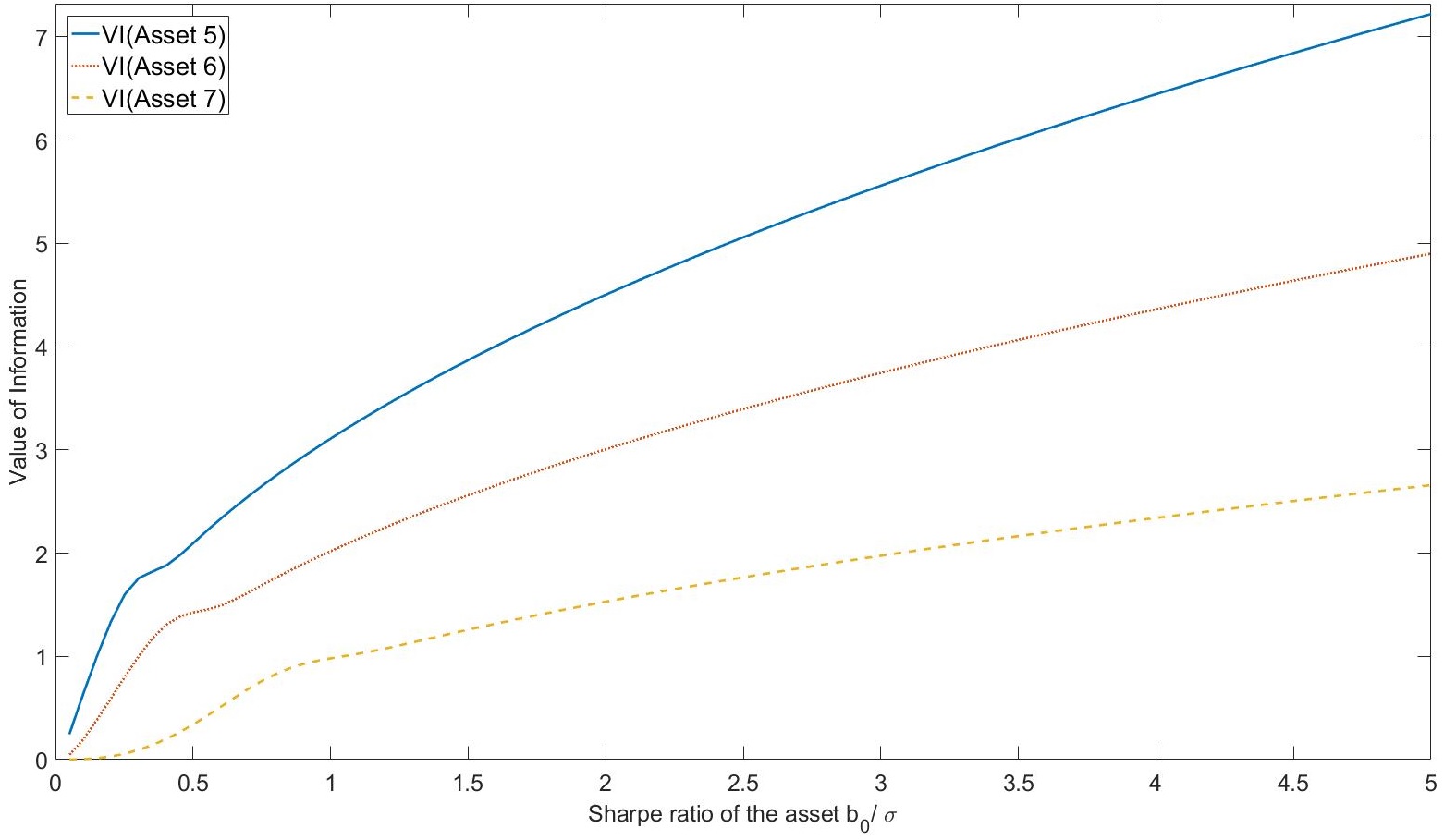}
  \caption{Value of information as a function of the Sharpe ratio of the asset with parameters as in Table \ref{tab: F3 parameters}.}
  \label{graph: Sharpe_ratio}
\end{figure}

\begin{table}[H]     
      \centering 
      \caption{Parameter values used in Figure \ref{graph: Sharpe_ratio}.}
      \begin{tabular}{cccccc} 
        \cline{2-6}
          \multicolumn{1}{c}{}&$\mathbf{b_0}$ & $\boldsymbol{\sigma}$ & \textbf{Sharpe ratio} & \textbf{T} &  $\boldsymbol{\sigma}_\mathbf{0}$\\
         \hline
         $\mathbf{Asset\;5}$ &5\% & [1-100]\%& [0-5] & 1 & 75\% \\
        $\mathbf{Asset\;6}$ &5\% & [1-100]\% & [0-5] & 1 &  35\% \\
        $\mathbf{Asset\;7}$ &5\% &  [1-100]\%& [0-5] & 1 &  10\% \\
        \hline
      \end{tabular}
      \label{tab: F3 parameters}
\end{table}
Figure \ref{graph: Sharpe_ratio} exhibits the value of information as a function of the Sharpe ratio of the asset, defined as $b_0/ \sigma$, for a sample of three assets with parameters set in Table \ref{tab: F3 parameters}.

We see that the higher the Sharpe ratio of the asset the bigger the value of information for any level of $\sigma_0$. Obviously, the higher the volatility of the drift, the more value information has and the more necessary it is to update the strategy. We also notice that the slope of the curve is steeper when the Sharpe ratio of the asset is low, ranging roughly from $0$ to $1$. Intuitively, it shows that information has more value for assets with a low Sharpe ratio because assets with a high Sharpe ratio will perform well, no matter the learning. Another explanation is for a fix $b_0$, a low Sharpe ratio means a high volatility which increases the need for learning and consequently the value of information. Furthermore, we clearly see that the curve with the higher $\sigma_0$ dominates the lower one, which confirms the intuition and understanding of Figure \ref{graph: drift_volatility}.

\subsubsection{Time} \label{subsec: time}

\begin{figure}[H] 
	\centering
  \includegraphics[scale=0.25]{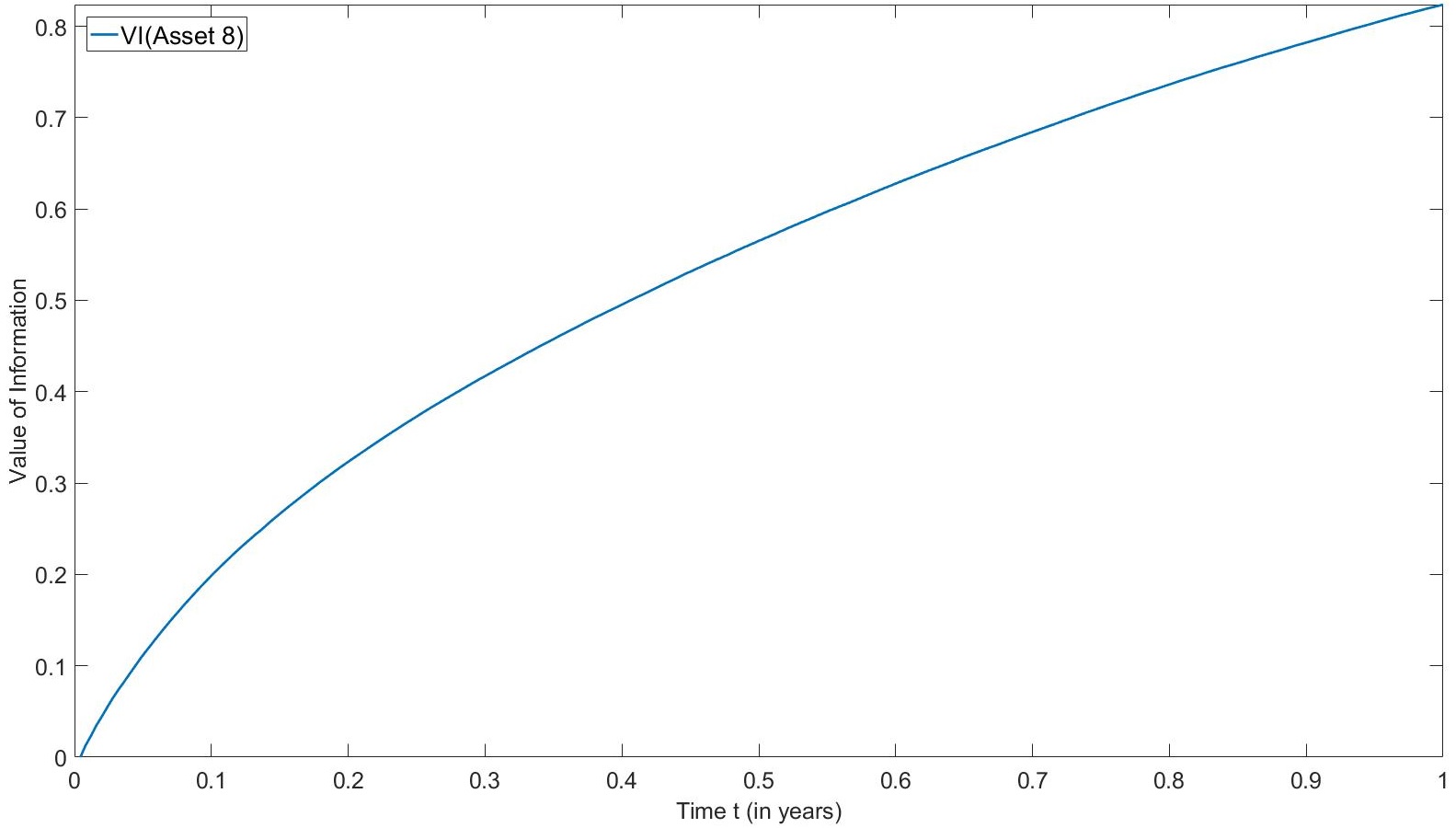}  
  \caption{Value of information as a function of t with parameters as in Table \ref{tab: F4 parameters }.}
  \label{graph: Sharpe_t}
\end{figure}

Figure \ref{graph: Sharpe_t} displays the value of information along the life of the investment for time $t\in [0,T]$. 
To obtain this graph, we use the set of parameters shown in Table \ref{tab: F4 parameters }. We simulate $N=1,000,000$ optimal wealth trajectories $(X)^i_{i\in [\![1,N]\!]}$ and at each point in time we compute the empirical Sharpe ratio of the learning strategy:

\beqs
\hat{Sh}^L_t &=&  \frac{\sqrt{N-1}}{N}\frac{\sum_{i=1}^N (X^i_t - x_0) }{\sqrt{\sum_{i=1}^N (X^i_t)^2 - \left( \sum_{i=1}^N X^i_t \right) ^2}}
\enqs
and use Formula \eqref{VXNL} for the non-learning strategy.

    \begin{table}[H]
      \centering 
      \caption{Parameter values used in Figure \ref{graph: Sharpe_t}}
      \begin{tabular}{ccccccc} 
        \cline{2-7}
        \multicolumn{1}{c}{}&$\mathbf{b_0}$ & $\boldsymbol{\sigma}$ & $\boldsymbol{\sigma_0}$ & \textbf{T} & $\boldsymbol{\vartheta}$ & \textbf{Simulations}\\
        \hline
        \textbf{Asset 8} &5\% & 20\% & 40\% & 1 & $(10\%)^2$ &1,000,000\\
        \hline
      \end{tabular}
      \label{tab: F4 parameters }
    \end{table}

As Figure \ref{graph: Sharpe_t} shows, the value of information increases monotonically with time. Nonetheless, the speed of increase tends to slow as time goes by. The fact that the marginal gain on the value of information decreases with time is well known and analyzed in the recent article by \cite{Keppo2018} in the context of investment decisions and costs of data analytics.


\subsubsection{Investment horizon} \label{subsec: Invest_Horiz}

\begin{figure}[H] 
	\centering
  \includegraphics[scale=0.25]{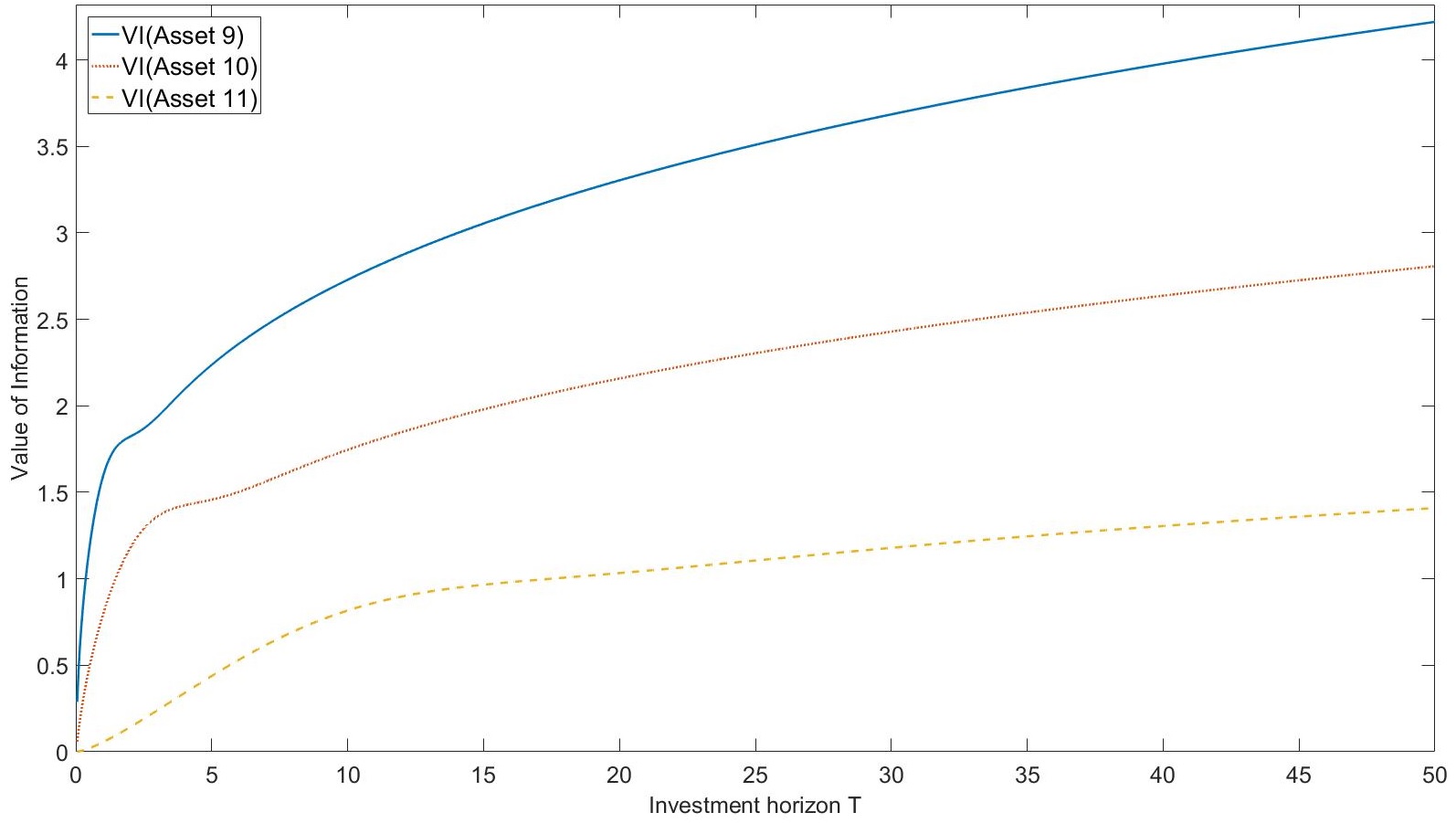}  
  \caption{Value of information as a function of T with parameters as in Table \ref{tab: F5 parameters}.}
  \label{graph: Invest_Horiz}
\end{figure}

Figure \ref{graph: Invest_Horiz} shows the value of information w.r.t. the investment time horizon $T$ for a sample of three assets with parameters described in Table \ref{tab: F5 parameters}.

\begin{table}[H]
      \centering 
      \caption{Parameter values used in Figure \ref{graph: Invest_Horiz}.}
      \begin{tabular}{cccccc} 
        \cline{2-6}
        \multicolumn{1}{c}{}&$\mathbf{b_0}$ & $\boldsymbol{\sigma}$ & \textbf{Sharpe ratio} & \textbf{T} &  $\boldsymbol{\sigma}_\mathbf{0}$\\
        \hline
        $\mathbf{Asset\;9}$ & 5\% & 20\% & 25\% & [0-50] & 75\% \\
        $\mathbf{Asset\;10}$ &5\% & 20\% & 25\% & [0-50] &  35\% \\
        $\mathbf{Asset\;11}$ & 5\% & 20\% & 25\% & [0-50] &  10\% \\
        \hline
      \end{tabular}
      \label{tab: F5 parameters}
\end{table}

Although the three curves are increasing no matter the value of the other parameters, the slope of these curves are mainly decreasing with the investment horizon. It suggests that the investor should consider that the marginal contribution to the value of information decreases with the investment horizon. It means that the spread between the optimal, Bayesian-Markowitz, and a sub-optimal, non-learning, strategy tightens for long investment horizons. It makes sense since for long horizons, at some point, the posterior distribution will not move so much since the values of the parameters of the distribution will be marginally affected by more learning. In addition, we remark that the bigger the drift volatility parameter, the higher the level of the associated value of information curve. This confirms the analysis of Figure \ref{graph: drift_volatility} and \ref{graph: Sharpe_ratio}. 


\section{Conclusion} \label{sec: Conclusion}

In this paper, we have solved the Bayesian-Markowitz problem when the unknown drift is assumed to follow a probability distribution corresponding to a prior belief.  
The investor then updates her information on the drift from a predictive distribution based on observed data coming from assets prices. 
We have turned the non-standard problem into a standard one to exhibit the HJB equation associated to the standardized problem and apply dynamic programming techniques. To illustrate our theoretical results, we have computed the key diffusion coefficient  $\psi$ in the case of the multidimensional discrete law and the Gaussian law, and provided  the full solution to the problem in the multidimensional Gaussian case. We have described a way of measuring the performance of investment strategies, the Sharpe ratio of terminal wealth, and used it to assess the value of information. To exhibit the value added of implementing a Bayesian learning approach in our problem, we have illustrated the value of information sensitivity to various key parameters and concluded that learning brings value to the optimal strategy that solves the Markowitz problem in a framework of drift uncertainty modeled, by a prior distribution with a positive definite covariance matrix, and a constant known volatility.

\section{Appendices}\label{ sec: append}
\appendix
\section{Proof of Lemma \ref{lem: dual_transform}} \label{sub:proof_lemma_duality}
\begin{proof}
	We use similar arguments as in \cite*{PhamIsmail2017}.
    \\
    
	Fix $\vartheta >0$, for any $\epsilon>0$, there exists an $\epsilon$-optimal control for $U_0(\vartheta)$, $\al^{\epsilon} \in \mathcal{A}$ such that 
	\beqs\label{}
		U_{0}(\vartheta) \leq \E[X^{\al^{\epsilon}}_T]+\epsilon \text{ and } \text{Var}(X^{\al^{\epsilon}}_T) \leq \vartheta.
	\enqs
	Then for any $\lambda >0$,
	\beqs
    \begin{aligned} 
 		V_{0}(\lambda) \leq \lambda \text{Var}(X^{\al^{\epsilon}}_T)-\E[X^{\al^{\epsilon}}_T] \leq \lambda \vartheta - U_{0}(\vartheta) +  \epsilon. 
	\end{aligned}
	\enqs   
	Since $\epsilon$ is arbitrary and the above relation holds for any $\vartheta$ we deduce,
    \beqs
	\begin{aligned}\label{eq: ineg V0}
 		V_{0}(\lambda)\leq \inf_{\vartheta >0} [\lambda \vartheta - U_{0}(\vartheta)],\quad \forall \lambda >0.
	\end{aligned}
    \enqs
	Conversely, fix $\lambda>0$ and $\epsilon>0$ and consider the $\epsilon$-optimal control $\al^{*,\lambda, \epsilon} \in \mathcal{A}$ for $V_0$ such that
	\beqs
		V_0(\lambda) \geq \lambda {\rm Var}(X_T^{\al^{*,\lambda,\epsilon}}) - \E[X_T^{ \al^{*, \lambda,\epsilon}}] - \epsilon.
	\enqs
	If we define
	\beqs
		\vartheta^{\lambda, \epsilon} : = {\rm Var}(X_T^{\al^{*, \lambda, \epsilon}}),
	\enqs
	by definition of $U_{0}(\vartheta^{\lambda, \epsilon})$ the inequality $ \E[X_T^{\al^{*, \lambda, \epsilon}}] \leq U_{0}(\vartheta^{\lambda, \epsilon})$ holds true.
    \\
    Hence,
	\begin{align*}
		V_0(\lambda) \geq &\lambda \vartheta^{\lambda, \epsilon}  - \E[X_T^{\al^{*, \lambda, \epsilon}} ] - \epsilon 
  		\geq  \lambda \vartheta^{\lambda, \epsilon}  - U_{0}(\vartheta^{\lambda, \epsilon})- \epsilon
	\end{align*}
	which holds for arbitrary $\epsilon>0$. This shows the first duality relation, i.e. $V_0$ is the Fenchel Legendre transform of $U_0$, and $\vartheta^\lambda$ attains the infimum, namely
\beqs
 		V_{0}(\lambda) = \inf_{\vartheta >0} [\lambda \vartheta - U_{0}(\vartheta)] = \lambda \vartheta^\lambda - U_{0}(\vartheta^\lambda).
\enqs
We now prove the second duality relation.
Fix $\vartheta>0$ and for any $\lambda>0$ consider the optimal control $\hat \al$ for $U_0$, we then have
\beqs
	\lambda \vartheta -U_0(\vartheta) \geq \lambda {\rm Var}(X^{\hat \al}_T) - \E[X^{\hat \al}_T] \geq V_0(\lambda).
\enqs
The previous relation holds for any $\lambda>0$, thus
\beq \label{eq: Maj_U_0}
	 U_0(\vartheta) \leq \inf_{\lambda>0} [\lambda \vartheta-V_0(\lambda)].
\enq
Now fix $\vartheta>0$ and consider $\al^{*,\lambda}$ the optimal control for $V_0$, we have
\beqs
	V_0(\lambda) = \lambda {\rm Var}(X_T^{\al^{*,\lambda}})-\E[X_T^{\al^{*,\lambda}}].
\enqs
Choose $\lambda(\vartheta$)  such that ${\rm Var}(X_T^{\al^{*,\lambda(\vartheta)}}) = \vartheta$. We then have
\beqs
	\begin{aligned}
    	V_0(\lambda(\vartheta)) &= \lambda(\vartheta) {\rm Var}(X_T^{\al^{*,\lambda(\vartheta)}})-\E[X_T^{\al^{*,\lambda(\vartheta)}}]\\
        &= \lambda(\vartheta) \vartheta-\E[X_T^{\al^{*,\lambda(\vartheta)}}]\\
        &\geq \lambda(\vartheta) \vartheta-U_0(\vartheta)
    \end{aligned}
\enqs
by definition of $U_0$. Together with \ref{eq: Maj_U_0}, this shows the second duality relation; $U_0$ is the Fenchel-Legendre transform of $V_0$ and $\lambda(\vartheta)$ attains the infimum in this transform
\beqs
	U_0(\vartheta) = \inf_{\lambda>0}[\lambda \vartheta - V_0(\lambda)]= \lambda(\vartheta) \vartheta-V_0(\lambda(\vartheta)).
\enqs
Finally, for $\lambda > 0$, if $\al^{*, \lambda} \in \mathcal{A}$ is the optimal control for $V_0$ i.e. for problem \eqref{eq: mean_variance_problem} then
	\beqs
 		V_{0}(\lambda) = \lambda {\rm Var}(X_T^{\al^{*,\lambda}}) - \E[X_T^{\al^{*, \lambda}}] = \lambda \vartheta^{\lambda} - U_{0}(\vartheta^\lambda)
	\enqs
with $\vartheta^{\lambda}:= {\rm Var}(X_T^{\al^{*,\lambda}})$. Moreover, we have

\beqs
	\begin{aligned}
    	U_0(\vartheta) = \inf_{\lambda>0} [\lambda \vartheta-V_0(\lambda)]&= \lambda(\vartheta) \vartheta-V_0(\lambda(\vartheta))\\
        &= \lambda(\vartheta) \vartheta-\lambda(\vartheta)\vartheta^{\lambda(\vartheta)} + \E[X_T^{\al^{*, \lambda(\vartheta)}}]\\
        &= \E[X_T^{\hat \al^{\vartheta}}]
    \end{aligned}
\enqs
since $\vartheta^{\lambda(\vartheta)}=\vartheta$, by definition of $\lambda(\vartheta)$. This means that $\hat \al^\vartheta$ is an optimal control for $U_0$ i.e. problem \eqref{eq: markowitz_problem}.
\end{proof}

\section{Proof of Lemma \ref{lem: quadratic_transform}}\label{sub:proof_lemma_quadratic_transform}
  \begin{proof}
    We use a similar approach as in \cite*{Zhou2000}.
    \\
   Since $\forall \al \in \mathcal{A}$, we can write
    \beqs
    	\E[X_T^{\al}]^{2} = - \inf_{\gamma\in\R}\left[ \gamma^{2}-2\gamma \E[X_T^{\al}]\right]
    \enqs
    where the infimum is achieved for $\gamma^* = \E[X_T^{\al}]$, we then have
    \begin{align*}
      V_0(\lambda) &= \inf_{\al \in \mathcal{A}}\left[\lambda \text{Var}(X_T^{\al}) - \E[X_T^{\al}]\right]\\
      &= \inf_{\al \in \mathcal{A}}\left[\lambda \big( \E[(X_T^{\al})^{2}]-\E[X_T^{\al}]^{2}\big )- \E[X_T^{\al}]\right]\\
      &= \inf_{\al \in \mathcal{A}}\left[\lambda \left( \E[(X_T^{\al})^{2}]+\inf_{\gamma\in\R} \left[\gamma^{2}-2\gamma \E[X_T^{\al}]\right]\right)- \E[X_T^{\al}]\right] \\
      &= \inf_{\al \in \mathcal{A}}\left[\inf_{\gamma\in\R}\left[\lambda \left( \E[(X_T^{\al})^{2}]+ \gamma^{2}-2\gamma \E[X_T^{\al}]\right )- \E[X_T^{\al}]\right]\right]  \\	
      &= \inf_{\gamma\in\R}\left[\inf_{\al \in \mathcal{A}}  \left[\lambda \E[(X_T^{\al})^{2}]-(1+2\lambda\gamma) \E[X_T^{\al}]\right] +\lambda \gamma^{2}\right]\\
      &= \inf_{\gamma\in\R}\left[ \tilde V_0 (\lambda,\gamma)+ \lambda \gamma^2\right],
    \end{align*}
    where $\tilde V_0 (\lambda,\gamma) := \inf_{\al \in \mathcal{A}}  \left[\lambda \E[(X_T^{\al})^{2}]-(1+2\lambda\gamma) \E[X_T^{\al}]\right]$ as in \reff{eq: V1_quadratic_def}, which proves \eqref{eq: V1_quadratic}. 
    \\
    
    The function $\tilde{V}_0$ is clearly linear in $\gamma$, so $\tilde V_0 (\lambda,\gamma)+ \lambda \gamma^2$ is strictly convex in $\gamma$ and the infimum $V_0(\lambda)$ exists for some unique $\gamma^*(\lambda) :=  {\rm argmin}_{\gamma \in \R} [\tilde V_0(\lambda, \gamma^*)+\lambda (\gamma^*)^2]$.
Take now this $\gamma^*(\lambda)$, $\atil^{\lambda, \gamma}$ an optimal control for $\tilde V_0(\lambda, \gamma)$ and $\alpha^{*,\lambda}  = \atil^{\lambda, \gamma^*(\lambda)}$.
    \\
    Then, dropping the dependence in $\lambda$ of $\gamma^*(\lambda)$ to alleviate notations, we have
    \beqs
    \underset{\gamma \in \R}{\rm argmin} [\tilde V_0(\lambda, \gamma^*)+\lambda (\gamma^*)^2] &= \underset{\gamma \in \R}{\rm argmin}\left[\lambda \E[(X^{\alpha^{*,\lambda}}_T)^{2}]-(1+2\lambda\gamma^*) \E[X^{\alpha^{*,\lambda}}_T]+\lambda (\gamma^*)^2\right]
    \enqs
    and find 
      \beqs
    	\gamma^* = \E[X^{\alpha^{*,\lambda}}_T],
    \enqs
 simply by differentiating the strictly convex function $ \gamma^* \mapsto \lambda \E[(X^{\alpha^{*,\lambda}}_T)^{2}]-(1+2\lambda\gamma^*) \E[X^{\alpha^{*,\lambda}}_T]+\lambda (\gamma^*)^2$.\\     
 Consequently, we have
  \begin{align*}
 	V_0(\lambda) &= \tilde V_0(\lambda,\gamma^*)+\lambda (\gamma^*)^2\\
    & = \lambda \E[(X^{\alpha^{*,\lambda}}_T)^{2}]-(1+2\lambda\gamma^*) \E[X^{\alpha^{*,\lambda}}_T]+\lambda (\gamma^*)^2\\
    &= \lambda \left( \E[(X^{\alpha^{*,\lambda}}_T)^{2}]-2\gamma^*\E[X^{\alpha^{*,\lambda}}_T]+(\gamma^*)^2\right)-\E[X^{\alpha^{*,\lambda}}_T]\\
    &=\lambda {\rm Var}(X^{\alpha^{*,\lambda}}_T)-\E[X^{\alpha^{*,\lambda}}_T]
 \end{align*}
 which shows that $\al^{*,\lambda}$ is an optimal control for $V_0$. 
  \end{proof}
 
 \section{Proof of theorem \ref{th: theomain}} \label{proof prop: th: main}
 \begin{proof}
   Following the standard dynamic programming approach as in \cite*{Pham2009}, we derive the HJB equation for the standard control problem \reff{eq: value function}:
   \beq \label{eq: HJB}
     \left \{
     \begin{aligned}
       & \dt v^{\lambda, \gamma} + \frac{1}{2} \text{tr} \left( \psi \psi^\trp \Hessb v^{\lambda, \gamma} \right) + \inf_{\al \in \mathcal{A}} \left \{ \alt^\trp  b \dx v^{\lambda, \gamma} +\frac{1}{2}|\sigma^\trp \alt|^2  \dxx v^{\lambda, \gamma}  + \alt^\trp \sigma \psi^\trp  \dxb v^{\lambda, \gamma} \right \}  = 0, \\
       & v^{\lambda, \gamma}(T,x,b) =  \lambda x^{2} -(1+2\lambda \gamma) x,
     \end{aligned}
     \right.
   \enq  
   where the matrix function $\psi(t,b)$ is noted $\psi$ to alleviate notations.
   \\
   
   Assuming for now that $\dxx v^{\lambda,\gamma} \geq0$, we find 
   \begin{equation}\label{eq: optimal a}
   	\underset{\al \in \mathcal{A}}{\rm argmin} \left \{ \alt^\trp  b \dx v^{\lambda, \gamma} +\frac{1}{2}|\sigma^\trp \alt|^2  \dxx v^{\lambda, \gamma}  + \alt^\trp \sigma \psi^\trp  \dxb v^{\lambda, \gamma} \right \}=-\Sigma^{-1} b \frac{\dx v^{\lambda, \gamma}}{\dxx v^{\lambda, \gamma}}-(\psi \sigma^{-1})^\trp \frac{\dxb v^{\lambda, \gamma}}{\dxx v^{\lambda, \gamma}}
   \end{equation}
   which turns the standard control problem \reff{eq: HJB} into the same following problem
   \beq\label{eq: problem_with_a_opt}
     \left \{
     \begin{aligned}
       & \dt v^{\lambda, \gamma}+ \frac{1}{2} \text{tr}\left( \psi \psi^\trp \Hessb v^{\lambda, \gamma} \right) - \frac{1}{2 \dxx v^{\lambda, \gamma}} |\dx v^{\lambda, \gamma} \sigma^{-1} b+\psi^\trp \dxb v^{\lambda, \gamma}|^{2}  = 0 \\
       & v^{\lambda, \gamma}(T,x,b) =  \lambda x^{2} -(1+2\lambda \gamma)x.
     \end{aligned}
     \right.
   \enq
   We look for a solution in the ansatz form $v^{\lambda, \gamma}(t,x,b) = K(t,b)x^{2}+\Gamma(t,b)x+ \chi(t,b)$ with $K \geq 0$ which ensures $\dxx v^{\lambda,\gamma} \geq 0$ and \reff{eq: optimal a}. Formally we derive,
   \beqs
     \begin{aligned}
       \dt v^{\lambda, \gamma}  =& \dt K x^2+ \dt \Gamma x + \dt \chi, &&&&
       \dx v^{\lambda, \gamma}   = 2Kx+\Gamma, \\
       \dxx v^{\lambda, \gamma}  =& 2K, &&&&
       \gradb v^{\lambda, \gamma}  = \gradb K x^2+ \gradb \Gamma x + \gradb \chi, \\
       \Hessb v^{\lambda, \gamma}  =& \Hessb K x^2+ \Hessb \Gamma x + \Hessb \chi,&&&&
       \dxb v^{\lambda, \gamma}   = 2 \gradb K x+ \gradb \Gamma.\\
     \end{aligned}
   \enqs
   Plugging the previous partial derivatives into \eqref{eq: problem_with_a_opt}, 
  the problem becomes:
   \beqs
     \left \{
     \begin{aligned}
       &0 = \left( \dt K + \frac{1}{2} \text{tr} \left( \psi \psi^\trp \Hessb K \right)-\left ( 2(\psi \sigma^{-1}b)^\trp \gradb K + \frac{|\psi^\trp \gradb K |^2}{K} +|\sigma^{-1}b|^2 K \right) \right )x^2\\
       &+ \left( \dt \Gamma+\frac{1}{2}\text{tr} \left( \psi \psi^\trp \Hessb \Gamma \right)-\left( \left( (\psi \sigma^{-1}b)^\trp +\frac{ \left( \psi \psi^\trp \gradb K \right)^\trp }{K} \right)\gradb \Gamma  + \left(|\sigma^{-1}b|^2 + \left(\psi \sigma^{-1}b \right)^\trp \frac{\gradb K}{K}  \right)\Gamma\right) \right)x\\
       &+ \dt \chi + \frac{1}{2} \text{tr} \left( \psi \psi^\trp \Hessb \chi \right) - \frac{1}{4K}|\sigma^{-1}b \Gamma + \psi^\trp \gradb \Gamma|^2,\\
       &v(T,x,b) =  \lambda x^{2} -(1+2\lambda \gamma)x,
     \end{aligned}
     \right.
   \enqs
   and by identification we obtain the following system of PDEs:
   \beqs
     \left \{
     \begin{aligned}
       0 = & \dt K + \frac{1}{2} \text{tr}\left(  \psi \psi^\trp \Hessb K \right)-\frac{1}{K} |\sigma^{-1}b K + \psi^\trp \gradb K|^2,   \\
       0 = & \dt \Gamma+\frac{1}{2}\text{tr} \left(  \psi \psi^\trp\Hessb \Gamma \right)-\left( \left((\psi \sigma^{-1}b)^\trp +\frac{ ( \psi \psi^\trp \gradb K)^\trp }{K} \right)\gradb \Gamma + \left(|\sigma^{-1}b|^2 + \left( \psi \sigma^{-1}b \right)^\trp \frac{\gradb K}{K}  \right)\Gamma\right), \\ 
       0 = & \dt \chi + \frac{1}{2} \text{tr} \left(  \psi \psi^\trp \Hessb \chi \right) - \frac{1}{4K}|\sigma^{-1}b \Gamma + \psi^\trp \gradb \Gamma|^2,
     \end{aligned}
     \right.
   \enqs
   with terminal conditions
   \beqs
     \left \{
     \begin{aligned}
       & K(T,b) = \lambda, \\
       & \Gamma (T,b) = -(1+2\lambda \gamma), \\
       & \chi (T,b) = 0.\\
     \end{aligned}
     \right.
   \enqs
   We now introduce the functions $K = \lambda\Ktil  $, $\Gamma = -(1+2 \lambda \gamma) \Gamtil$ and $\chi = (1+2 \lambda \gamma)^2 \Xitil/\lambda$, so that the previous system becomes
   \beqs
     \left \{
     \begin{aligned}
       0=& \dt \Ktil + \frac{1}{2} \text{tr}\left( \psi \psi^\trp \Hessb \Ktil  \right)-\frac{1}{\Ktil}|\sigma^{-1}b \Ktil + \psi^\trp \gradb \Ktil|^2,   \\
       0 = & \dt \Gamtil+\frac{1}{2}\text{tr} \left( \psi \psi^\trp \Hessb \Gamtil  \right)-\left( \left( \left( \psi \sigma^{-1}b \right )^\trp +\frac{ ( \psi \psi^\trp \gradb \Ktil)^\trp }{\Ktil} \right) \gradb \Gamtil + \left(|\sigma^{-1}b|^2 + (\psi \sigma^{-1}b)^\trp \frac{\gradb \Ktil}{\Ktil}  \right)\Gamtil\right),  \\ 
       0=& \dt \Xitil + \frac{1}{2} \text{tr} \left( \psi \psi^\trp \Hessb \Xitil \right) - \frac{1}{4 \Ktil}|\sigma^{-1}b \Gamtil+ \psi^\trp \gradb \Gamtil|^2,
     \end{aligned}
     \right.
   \enqs
   with terminal conditions
   \beqs
     \left \{
     \begin{aligned}
       & \Ktil(T,b) = 1, \\
       & \Gamtil (T,b) = 1, \\
       & \Xitil (T,b) = 0. \\
     \end{aligned}
     \right.
   \enqs
   If there exists a unique solution of the equation for $\Ktil$, then $\Gamtil = \Ktil$ is the unique solution that verifies the linear PDE for $ \Gamtil$ and $\Xitil = (\Ktil-1)/4$ is the unique solution of the linear PDE for $\Xitil$. 
\\

Since we look for positive $\Ktil$, we introduce $\Ktil(t,b):= e^{-R(t,b)}$ and rewrite the related PDE:
   \beqs
     \left \{
     \begin{aligned}
       0 = &\Ktil\left(- \dt R - \frac{1}{2} \text{tr}\left( \psi \psi^\trp \Hessb R  - \psi \psi^\trp \gradb R\gradb R^\trp  \right) + 2 \left(\psi \sigma^{-1}b \right)^\trp \gradb R  - |\psi^\trp  (\gradb R)|^2  -|\sigma^{-1}b|^2 \right), \\
       0=& R(T,b). \\
     \end{aligned}
     \right.
   \enqs
   From $\text{tr}\left(\psi \psi^\trp \gradb R \gradb R^\trp  \right) =  |\psi^\trp  \gradb R|^2$ we finally obtain the expression of the semi-linear PDE in \reff{eq: PDE_w}:
   \beqs
     \left \{
     \begin{aligned}
       0 = & - \dt R - \frac{1}{2} \text{tr}\left( \psi \psi^\trp \Hessb R \right)+ 2 \left(\psi \sigma^{-1}b \right)^\trp \gradb R - \frac{1}{2}|\psi^\trp (\gradb R)|^2 -|\sigma^{-1}b|^2,\\
       0 = & R(T,b) .
     \end{aligned}
     \right.
   \enqs
We obtain the following value function dependent upon $(\lambda, \gamma)$:
  \begin{align} \label{eq: v_l_g}
    \vtxbgl = &  \lambda \Ktil(t,b)x^{2}-(1+2\lambda \gamma)\Ktil(t,b) x+\frac{(1+2\lambda \gamma)^2}{\lambda} \frac{\Ktil(t,b)-1}{4} \nonumber \\ 
    = & e^{-R(t,b)}\left(\lambda x^{2}-(1+2\lambda \gamma) x+\frac{(1+2\lambda \gamma)^2}{4\lambda}\right) -\frac{(1+2\lambda \gamma)^2}{4\lambda},
  \end{align}
  and from Eq. \eqref{eq: optimal a}, the optimal feedback control $\tilde{\al}^{\lambda, \gamma}$ is:
  \beqs \label{eq: a_l_g}
  	\tilde{\al}^{\lambda, \gamma}=\tilde{a}^{\lambda, \gamma}(t,X^{\tilde{\al}^{\lambda, \gamma}}_t,b) ,
  \enqs
  with
  \beqs 
  	\tilde{a}^{\lambda, \gamma}(t,x,b) = \left(\frac{1}{2 \lambda}+\gamma-x\right) \left( \Sigma^{-1} b - (\psi \sigma^{-1}).^\trp \gradb R(t,b)\right).
  \enqs
Moreover, from $\tilde{V_0}(\lambda, \gamma) = v^{\lambda,\gamma}(0, x_0, b_0) $ we compute
  \begin{align*}\label{eq: optimal gamma}
    \gamma^*\left(\lambda\right) = & \underset{\gamma \in \R}{\rm argmin} \left[ \tilde{V_0}(\lambda, \gamma)  + \lambda \gamma^2\right] \\ 
    = &  \underset{\gamma \in \R}{\rm argmin} \left[ e^{-R(0,b_0)}\left(\lambda x_0^{2}-(1+2\lambda \gamma) x_0+\frac{(1+2\lambda \gamma)^2}{4\lambda}\right) -\frac{(1+2\lambda \gamma)^2}{4\lambda}  + \lambda \gamma^2\right] \\ 
    = &  \underset{\gamma \in \R}{\rm argmin} \left[ e^{-R(0,b_0)}\left(\lambda x_0^{2}-x_0+\frac{1}{4\lambda}\right)  -\frac{1}{4\lambda}  + \left( (1-2\lambda x_0) e^{-R(0,b_0)}-1 \right)\gamma  + \lambda  e^{-R(0,b_0)} \gamma^2 \right] \\ 
    = & x_0  + \frac{1}{2\lambda} \left(e^{R(0,b_0)} -1\right). \\
  \end{align*}
  Remembering \eqref{eq: v_l_g}, the value function at time 0 with the optimal $\gamma$ is computed as followed:
  \beqs
    \begin{aligned}
      v^{\lambda, \gamma^*}(0,x_0,b_0)=& e^{-R(0,b_0)}\left(\lambda x_0^{2}-(2\lambda x_0 +e^{R(0,b_0)} x_0) \right)+ \left( \lambda x_0^2 + \frac{e^{2R(0,b_0)}}{4 \lambda} +x_0 e^{R(0,b_0)}\right) \left(e^{-R(0,b_0)}-1 \right) \\
      =& -\lambda x_0^2 + \frac{e^{R(0,b_0)}}{4 \lambda} \left(1-e^{R(0,b_0)} \right) - x_0e^{R(0,b_0)},
    \end{aligned}
  \enqs
  and using Eq. \eqref{eq: V1_quadratic}, we find the expression in \reff{eq: V_0_lambda}:
  \beqs	
    \begin{aligned} 	
      V_0(\lambda) &= \tilde{V_0}(\lambda,\gamma^*) + \lambda (\gamma^*)^2 =  v^{\lambda, \gamma^*}(0,x_0,b_0) + \lambda (\gamma^*)^2 \\
      & = \frac{e^{R(0,b_0)}}{4 \lambda} \left(1-e^{R(0,b_0)} \right)- x_0e^{R(0,b_0)} +\lambda x_0^2 + \frac{e^{2R(0,b_0)}}{4 \lambda} +\frac{1}{4 \lambda} + x_0 e^{R(0,b_0)} -x_0 - \frac{e^{R(0,b_0)}}			{2\lambda}\\
      & = - \frac{1}{4 \lambda} \left( e^{R(0,b_0)}-1\right) -x_0.
    \end{aligned} 
  \enqs
\\
From Lemma \reff{lem: quadratic_transform}, we find the optimal terminal wealth in \reff{eq: optterminal}
\beqs
	\E[X^{\al^{*,\lambda}}_T] = \gamma^*(\lambda) =  x_0  + \frac{1}{2\lambda} \left(e^{R(0,b_0)} -1\right),
\enqs
obtained with the optimal feedback control:
\beqs
	\al_t^{*,\lambda} = \tilde a^{\lambda,\gamma^*(\lambda)}(t,X^{\al^{*,\lambda}}_t,\hat B_t) = a_0^{Bayes,\lambda}(t,X^{\al^{*,\lambda}}_t,\hat B_t),
\enqs
with 
\beq \label{eq:a_opt_lambda}
a_0^{Bayes, \lambda}(t,x,b) = \left(x_0-x+\frac{e^{R(0,b_0)}}{2 \lambda}\right) \left( \Sigma^{-1} b - (\psi \sigma^{-1})^\trp \gradb R(t,b) \right).
\enq
The Lagrange multiplier $\lambda(\vartheta)$ which makes the variance of the optimal wealth equals to the variance constraint $\vartheta$ is calculated as follows.
  We know that,
  \beq \label{eq: V_0}
  	V_0(\lambda(\vartheta)) = - \frac{1}{4 \lambda(\vartheta)} \left( e^{R(0,b_0)}-1\right) -x_0,
  \enq
and from Eq. \eqref{eq: mean_variance_problem}, we explicitly compute
  \beq\label{eq: V_0_lamb}
    \begin{aligned}
      V_0(\lambda(\vartheta))& = \lambda(\vartheta) {\rm Var} ( X^{\CM}_T) - \E [X^{\CM}_T]\\
      & = \lambda(\vartheta) \vartheta - \E [X^{\CM}_T] \\
      & = \lambda(\vartheta) \vartheta - x_0 - \frac{1}{2\lambda(\vartheta)}\left(e^{R(0,b_0)}-1 \right).\\ 	
    \end{aligned}
  \enq
  From Eq. \reff{eq: V_0} and \reff{eq: V_0_lamb} we obtain \reff{eq: explilambda}, namely
  \beq \label{eq: lambd_v}
  	\lambda(\vartheta) = \sqrt{\frac{1}{4\vartheta} \left( e^{R(0,b_0)}-1\right)}.
  \enq 
  From the correspondance of the controls established in Lemma \ref{lem: dual_transform}, we know that
  \beqs
  	\CM = \al ^{*, \lambda(\vartheta)}.
  \enqs
  We then compute the optimal controls of the Bayesian-Markowitz problem $U_0$ as
  \beqs
  	\CM=a_0^{Bayes,\lambda(\vartheta)}(t,X^{\CM}_t,\hat{B}_t),
  \enqs
 where, from Eq. \eqref{eq:a_opt_lambda} with $\lambda = \lambda(\vartheta)$ as in \eqref{eq: lambd_v},
  \beqs
  	a_0^{Bayes,\lambda(\vartheta)}(t,x,b) \;=\; \left(x_0-x + e^{R(0,b_0)}\sqrt{\frac{\vartheta}{e^{R(0,b_0)}-1}}\right) \left( \Sigma^{-1} b - (\psi \sigma^{-1})^\trp \gradb R(t,b)\right).
  \enqs
 Finally from \reff{eq: V_0_lamb} and \reff{eq: lambd_v}, we deduce the optimal performance of the Bayesian-Markowitz problem:
 \beqs
	\begin{aligned}
    	U_0(\vartheta) \; &=\; \lambda(\vartheta)\vartheta - V_0(\lambda(\vartheta))\\
        \; &=\; x_0 + \frac{1}{2\lambda(\vartheta)}\left(e^{R(0,b_0)}-1 \right)
        \; &=\; x_0 + \sqrt{\vartheta(e^{R(0,b_0)}-1)}.
    \end{aligned}
\enqs  
\end{proof}

\section{Proofs of theorem \ref{th: Existence Solution}} \label{proof: th: Existence Solution}
Without loss of generality, we consider in this demonstration the function $\tilde{R}:=-R$. The value of the positive constant $C$ may change from line to line and the matrix function $\psi(t,b)$ will be simply noted $\psi$ to alleviate notations.
\\

We define the nonlinear function $F:\mathcal{R} \times \Rn \rightarrow \R$ by:
\beqs
	 F(t,b,p) \;=\; \frac{1}{2}|\psi^\trp p|^2 + 2(\psi \sigma^{-1}b)^\trp p + |\sigma^{-1} b|^2,
\enqs
and we introduce the function $L_t: \mathcal{B}_t \times \Rn \rightarrow \R$ by
\beqs
	 L_t(b,q) \; := \; \max_{p\in \Rn} \big[  -q^\trp \psi^\trp p-F(t,b,p) \big].
\enqs
We notice that $F$ is quadratic hence convex in $p$, so it is easy to see that for fixed $(t,b) \in {\cal R}$ the function $p \mapsto -q^\trp \psi^\trp p-F(t,b,p)$ reaches a maximum $p^*= -(\psi^\trp)^{-1} (q + 2 \sigma^{-1} b)$. This gives us the explicit form of the function $L_t$ which depends on $t$ only through $\mathcal{B}_t$:
      \beqs
      	\begin{aligned}
     	 L_t(b,q) &= -q^\trp \psi^\trp p^*-F(t,b,p^*)\\
         &=\frac{1}{2}|q|^2 + 2(\sigma^{-1}b)^\trp q + |\sigma^{-1} b|^2.
      	\end{aligned}
      \enqs
      Conversely, the function $L_t$ is convex in $q$ so for fixed $b \in {\cal B}_t$ the function $q \mapsto -q^\trp \psi^\trp p-L_t(b,q)$ reaches its maximum for $q^* =- \left(\psi^\trp p + 2 \sigma^{-1}b \right)$. It shows that
      \beqs
        \begin{aligned}
          \max_{q \in \Rn} \big[  -q^\trp \psi^\trp p-L_t(b,q) \big] \; &=\; -(q^*)^\trp \psi^\trp p-L_t(b,q^*) \\
         \; &=\;\frac{1}{2}|\psi^\trp p|^2 + 2(\psi \sigma^{-1}b)^\trp p + |\sigma^{-1} b|^2 \\
         \;&=\;F(t,b,p),
        \end{aligned}
      \enqs
and it establishes the duality relation between the functions F and L.
Let us know consider the truncated function $F^k: {\cal R} \times \Rn \rightarrow \R$ defined for each $k \in \N$ by
\beqs 
	F^k(t,b,p) \;=\; \max_{q \in \mathcal{A}_k} \big[  -q^\trp \psi^\trp p-L_t(b,q) \big].
\enqs 
\\
      We observe from the explicit form of $L_t$ that $L_t \in C^{1}(\mathcal{B}_t \times \Rn)$ and $L$, $\gradb L$ satisfy a polynomial growth condition in $b$. Namely, we see that the following estimates hold true for $(b,q)\in \mathcal{B}_t \times \Rn$:
      \beqs\label{eq: L_estimate}
        \begin{aligned}
          |L_t(b,q)| \; &\leq \; \frac{1}{2}|q|^2 + 2|\sigma^{-1}||b||q| +|\sigma^{-1}|^2 |b|^2 \\
         \; & \leq \; \frac{1}{2}|q|^2 + |\sigma^{-1}|(|q|^2 + |b|^2) +|\sigma^{-1}|^2 |b|^2 \\
         \; & \leq \; C(|q|^2+|b|^2) 
        \end{aligned}
      \enqs
      and
      \beqs
        \begin{aligned}\label{eq: L_b_estimate}
          |\gradb L_t(b,q)|\; &\leq \; 2 ( |(\sigma^{-1})| |q| + |\Sigma^{-1}| |b|) \\
         \; &  \leq \; C (|q| + |b|). 
        \end{aligned}
      \enqs   
      So, $|L|$ and $|\gradb L|$  are of polynomial growth in $b$ uniformly in $|q|$ when $|q| \leq k$. By classical theory (see Theorem 4.3 p 163 in \cite*{Fleming2006}), the previous estimates and the assumptions of the theorem tell us that there exists a unique quadratically growing smooth solution 
      \beqs 
      	\tilde{R}^k \in C^{1,2} \left([0,T) \times {\cal B}_t \right) \cap C({\cal R}),
      \enqs 
to the truncated semi-linear PDE, 
\beqs 
	 - \dt{\Rtil^k} - \frac{1}{2}\text{tr}( \psi\psi^\trp \Hessb \Rtil^k) + F^k(t,b,\gradb \Rtil^k) \;=\;  0,
\enqs 
with terminal condition $\tilde{R}^k(T,.)=0$.
\\

      We then know from the Feynman-Kac formula and standard arguments that $\Rtil^k$ can be represented as the solution of the stochastic control problem
      \beq \label{eq: cont_pb}
      	\Rtil^k(t,b) = \inf_{q \in \mathcal{A}_k} \E \left[ \int_t^\trp L_t(\tilde{B}_s,q_s)ds |\tilde{B}_t = b\right]
      \enq
      where $\mathcal{A}_k$ is the compact set $\mathcal{A}_k=\{q \in \Rn : |q| \leq k \}, k > 0 $, and the dynamic of $\tilde{B}$ is
      \beq \label{eq: SDE}
     	 d\tilde{B}_s = \psi(s,\tilde{B}_s) \left(q_sds+dW_s\right).
      \enq
      Moreover, an optimal control for \eqref{eq: cont_pb} is Markovian and given by
      \beqs\label{eq: q_bounded}
      	q^{*}_k(t,b) = \underset{q \in \mathcal{A}_k}{\rm argmin} \left \{ q^\trp \psi^\trp \gradb \tilde{R}^k(t,b) + L_t(b,q)\right\}.
      \enqs
      We then deduce that, 
      \beqs \label{eq: cont_pb_alt}
      	\Rtil^k(t,b) = \E \left[ \int_t^\trp L_s \left(\tilde{B}_s^{*},q^{*}_k \left( s,\tilde{B}_s^* \right) \right) ds |\tilde{B}_t^* = b\right]
      \enqs
      where $\tilde{B}_s^*$ solves the stochastic differential equation (SDE) \eqref{eq: SDE} with controls $q^{*}_k \left( s,\tilde{B}_s^* \right)$.
      \\
      From the theorem of differentiation under the expectation and the integral sign we know that
      \beqs
      	\gradb \Rtil^k(t,b) = \E \left[ \int_t^\trp \gradb L_s \left(\tilde{B}_s^{*},q^{*}_k \left( s,\tilde{B}_s^* \right) \right) ds |\tilde{B}_t^* = b\right].
      \enqs
      \\
     To prove the linear growth condition of $\gradb \tilde{R}^k$, we will need the following inequality:
      \beqs
      	|\gradb L_t(b,q)|^2 \leq C_1 L_t(b,q) + C_2|b|^2,
      \enqs
      with $C_1 >0$ and $C_2 \geq 2 C_1|\sigma^{-1}b|^2>0$. 
      \\
      Having in mind that the function $L_t$ can be written as $L_t(b,q) = \frac{1}{2}|q+2\sigma^{-1}|^2-|\sigma^{-1}b|^2$, the previous inequality comes from:
      \beqs
        \begin{aligned}
          | \gradb L_t(b,q)|^2 \;&=\; |2 (\sigma^{-1})^\trp q + 2\Sigma^{-1}b|^2 \\
          \; & \leq \; 4 |\sigma^{-1}|^2 |q + 2\sigma^{-1}b-\sigma^{-1}b|^2 \\ 
          \; & \leq \; 8 |\sigma^{-1}|^2 \left( \frac{1}{2}|q + 2\sigma^{-1}b|^2+|\sigma^{-1}b|^2 \right)\\
          \; & \leq \; C_1 L_t(b,q) +2C_1|\sigma^{-1}|^2|b|^2 \\ 
          \; & \leq \; C_1 L_t(b,q) + C_2|b|^2. \\
        \end{aligned}
      \enqs
      Then by the Cauchy-Schwartz inequality and the fact that $L_t(b,0) \leq C|b|^2$ for some positive constant C independent of $k$ and $t$, we have
      \beqs
        \begin{aligned}
          |\gradb \Rtil^k(t,b)| &\leq  \E \left[ \int_t^T \Lnorm \gradb L_s \left(\tilde{B}_s^{*},q^{*}_k \left( s,\tilde{B}_s^* \right) \right) \Rnorm ds |\tilde{B}_t^* = b\right]\\
          & \leq C \E \left[ \int_t^T \Lnorm \gradb L_s \left(\tilde{B}_s^{*},q^{*}_k \left( s,\tilde{B}_s^* \right) \right) \Rnorm ^2ds |\tilde{B}_t^* = b\right]^{\frac{1}{2}}\\
          & \leq C \left( C_1 \E \left[ \int_t^T L_s\left(\tilde{B}_s^{*},q^{*}_k \left( s,\tilde{B}_s^* \right) \right)ds |\tilde{B}_t^* = b\right]+ C_2 \E \left[ \int_t^T |\tilde{B}_s|^2ds |\tilde{B}_t = b\right] \right)^{\frac{1}{2}}\\
          & \leq C \left( C_1 \E \left[ \int_t^T L_s\left(\tilde{B}_s,0 \right)ds |\tilde{B}_t = b\right] + C_2\E \left[ \int_t^T |\tilde{B}_s|^2ds |\tilde{B}_t = b\right]\right)^{\frac{1}{2}}\\
          & \leq C \E \left[ \int_t^T |\tilde{B}_s|^2ds |\tilde{B}_t = b\right]^{\frac{1}{2}}\\
          & \leq C(1+ |b|).
        \end{aligned}
      \enqs
      We used that when $q:=0$, $\tilde{B}_s = \tilde{B}_t + \int_t^s \psi(u, \tilde{B}_u) dW_u$ and thanks to Ito's formula,
      \beqs
      	|\tilde{B_s}|^2 =  |\tilde{B}_t|^2 + 2\tilde{B}_t^\trp\int_t^s \psi(u, \tilde{B}_u) dW_u + \int_t^s \text{tr} \left(\psi \psi^\trp(u, \tilde{B}_u) \right) du.
      \enqs
      Moreover,
      \beqs
        \begin{aligned}
          \E \left[ \int_t^\trp |\tilde{B}_s|^2ds |\tilde{B}_t = b\right]^{\frac{1}{2}} &=
          \E \left[ \int_t^\trp \left( |\tilde{B}_t|^2 + 2\tilde{B}_t^\trp\int_t^s \psi(u, \tilde{B}_u) dW_u + \int_t^s \text{tr}\left( \psi \psi^\trp(u, \tilde{B}_u)\right) du \right ) ds |\tilde{B}_t = b\right]^{\frac{1}{2}}\\
          & \leq \left ((T-t)|b|^2 + n|\psi|^2 (T-t) \right)^{\frac{1}{2}} \\
          & \leq C(1 + |b|^2)^{\frac{1}{2}}\\
          & \leq C \left(1 + |b| \right)  \\
        \end{aligned}
      \enqs
      \\
         Consequently, since the function $q \rightarrow  -q^\trp \psi^\trp \gradb R^k(t,b) - L_t(b,q)$ attains its maximum on $\Rn$ for
      \beqs\label{eq: qglobal}
      	\tilde{q}_k(t,b) = - \left( \psi^\trp \gradb R^k(t,b) + 2 \sigma^{-1}b    \right),
      \enqs
      and knowing that by assumption $|\psi| < \infty$, it is easy to see that
      \beqs \label{eq: qglob}
      	|\tilde{q}_k(t,b)| \leq C(1 + |b|).
      \enqs
      \\
      As a consequence, for an arbitrarily large constant $\hat{C}>0$, there exists a positive constant C independent of $k$ such that, 
      \beqs
      	|\tilde{q}_k(t,b)| \leq C, \quad t \in [0,T], \quad b \in \mathcal{B}_t \cap \{ |b| \leq \tilde{C} \}.
      \enqs
      \\
      Hence, for $k \geq C$, we have
      \beqs 
      	\begin{aligned}
			F^k(t,b,\gradb \tilde{R}^k) &= 	\max_{q \in \mathcal{A}_k} \big[  -q^\trp \psi^\trp \gradb \tilde{R}^k-L_t(b,q) \big], \\
            &= \max_{q \in \Rn} \big[  -q^\trp \psi^\trp \gradb \tilde{R}^k-L_t(b,q) \big], \\
            &= F(t,b,\gradb \tilde{R}^k).
        \end{aligned}
      \enqs 
      for all $(t,b) \in [0,T] \times \mathcal{B}_t \cap \{|b| \leq \tilde{C}\}$. Letting $\tilde{C}$ goes to infinity implies that for $k$ sufficiently large, $R^k = -\tilde{R}^k$ is a smooth solution satisfying a quadratic growth condition to \eqref{eq: PDE_w} - \eqref{eq: termcond}.

\section{Proof of Lemma \ref{lem: sol Gaussian Multi}} \label{proof lem: sol Gaussian Multi}
  \begin{proof}
    To solve the system in the multidimensional Gaussian case, we solve the Riccati equation for M. We look for a symmetric solution M such that $GM = (GM)^\trp$ which solves the following ODE
    \beqs
   	 -M'(t) -2M(t)^\trp G(t)^\trp \Sigma G(t) M(t) + 2G(t) M(t) + 2M(t)^\trp G(t)^\trp - \Sigma^{-1} =0. 
    \enqs
    A particular solution to the ODE for M is $\hat{M}(t) = G^{-1}(t)\Sigma^{-1}$ and it is easy to see that $\hat{M}$ is symmetric since $\Sigma G= G^\trp \Sigma$, and $G\hat{M} = (G\hat{M})^\trp$.
    \\
    
    Now we look for a function $N$ such that $M = \hat{M}+N$. Note that $N$ should be symmetric with $GN=(GN)^\trp$. Plugging $M$ into the ODE yields
    \beqs
    	-N'(t) - 2 N^\trp G(t)^\trp \Sigma G(t) N(t) = 0.
    \enqs
    Now, we change variable $N = \Theta^{-1}$ in the previous ODE to find
    \beqs
    	\Theta'(t) = 2G(t)^\trp \Sigma G(t).
    \enqs
    Noticing that $\Theta(T) = -\Sigma G(T) = -(\Sigma_0 ^{-1}+\Sigma^{-1} T)^{-1}$, we obtain after integration
    \beqs
    	\Theta(t) = -\left( (\Sigma_0 ^{-1}+\Sigma^{-1} T)^{-1}+\int_t^\trp2G(s)^\trp \Sigma G(s)ds \right).
    \enqs
    Finally, writing $G(t)$ as $ (\Sigma + \Sigma_0t)^{-1}\Sigma_0$, the solution to the original ODE is 
    \beqs
    	M(t)=(\hat{M}+\Theta^{-1})(t) =\Sigma_0 ^{-1}+\Sigma^{-1} t - \left[ (\Sigma_0 ^{-1}+\Sigma^{-1} T)^{-1}+2\int_t^\trp \Sigma_0 \left( \Sigma + \Sigma_0 s\right) ^{-1} \Sigma \left( \Sigma + \Sigma_0 s \right) ^{-1} \Sigma_0 ds \right ]^{-1}.
    \enqs
    To obtain $U(t)$, we simply integrate and with some simplifications we obtain,
    \beqs
    	U(t)= \int_t^T {\rm tr}\left( \Sigma_0 \left(\Sigma + \Sigma_0 s\right)^{-1} - \left[  \left( ( \Sigma_0^{-1}+ \Sigma^{-1}T) ^{-1} +2 \int_s^T G(u)^\trp \Sigma G(u)du\right) \left( G(s)^\trp \Sigma G(s) \right)^{-1} \right]^{-1}  \right)ds.
    \enqs
It is easy to verify that $M(T) = U(T) =0$.
\\

The next step is to check that N satisfies the conditions we have imposed to derive the solution $M$. The condition $N=N^\trp$ is satisfied since it is easy to see that $\Theta = \Theta^\trp$. It is straightforward to see that the second condition, $GN = (GN)^\trp$ is equivalent to $\Theta\Sigma_0^{-1} \Sigma = \Sigma \Sigma_0^{-1} \Theta$, once G is written as $(\Sigma + \Sigma_0t)^{-1}\Sigma_0$. Noticing the symmetry of $G \Sigma_0^{-1}$, we develop the left side of the equality,
\beqs
\begin{aligned}
	\Theta \Sigma_0^{-1} \Sigma &=  -\left( (\Sigma_0 ^{-1}+\Sigma^{-1} T)^{-1}+\int_t^\trp2G(s)^\trp \Sigma G(s)ds \right) \Sigma_0^{-1} \Sigma \\
    &= -\left( (\Sigma_0 ^{-1}+\Sigma^{-1} T)^{-1} \Sigma_0^{-1} \Sigma +\int_t^\trp2G(s)^\trp \Sigma G(s) \Sigma_0^{-1} \Sigma ds \right)\\
    &= -\left( \Sigma \left[(\Sigma \Sigma_0 ^{-1}+ T) \Sigma_0 \right]^{-1}  \Sigma +\int_t^\trp2 \Sigma G(s) \Sigma_0^{-1} G(s)^\trp \Sigma ds \right)\\
    &= -\left( \Sigma \Sigma_0 ^{-1} \left(\Sigma_0 ^{-1}+ \Sigma^{-1}T \right)^{-1}  +\int_t^\trp 2 \Sigma \Sigma_0^{-1} G^\trp(s)  \Sigma  G(s) ds \right)\\
    &= \Sigma \Sigma_0^{-1} \Theta.
\end{aligned}
\enqs

\end{proof}

\noindent For the one-dimensional case n = 1, we check that the functions:
    \beqs
    	M(t) = \frac{\sigma^2+\sigma_0^2 t}{\sigma^2 \left( \sigma^2+\sigma_0^2 (2T-t)\right)}(T-t) \quad \text{and} \quad U(t) = \text{log} \left ( \frac{\sigma^2+\sigma_0^2T}{\sqrt{(\sigma^2+\sigma_0^2t)(\sigma^2+	\sigma_0^2(2T-t))}}\right),
    \enqs		
    satisfy the following unidimensional ODE system:
    \beqs
    \left \{
    	\begin{aligned}
        	M'(t)  &= - \frac{2\sigma^2\sigma_0^4}{(\sigma^2+\sigma_0^2t)^2} M(t)^2 + \frac{4\sigma_0^2}{\sigma^2+\sigma_0^2t}M(t) - \sigma^{-2}, \\
            U'(t) &= -\frac{\sigma^2\sigma_0^4}{\sigma^2+\sigma_0^2t)^2}M(t),\\
        \end{aligned}
        \right.
    \enqs
    with M(T) = U(T) = 0.
    \\ 
    We easily see that M and U satisfy the terminal conditions. We first check that M satisfies the first ODE of the system. We derive the function $M$ w.r.t. $t$,
    \beqs		
   		M'(t) = \frac{\sigma^{-2}}{(\sigma^2+\sigma_0^2(2T-t))^2} \left(\sigma_0^2(-2\sigma^2t)+\sigma_0^4(2T^2-4Tt+t^2) - \sigma^4 \right),
    \enqs
    and notice that the right side of the equality for $M$ yields: 	
    \beqs
      \begin{aligned}
        &- \frac{2\sigma^2 \sigma_0^4}{(\sigma^2+\sigma_0^2t)^2}M(t)^2 + \frac{4 \sigma_0^2}{\sigma^2+\sigma_0^2t}M(t) - \sigma^{-2} \\
        & = \frac{\sigma^{-2}}{(\sigma^2+\sigma_0^2(2T-t))^2} \left( -2\sigma_0^4(T-t)^2+4\omega^2(T-t)(\sigma^2+\sigma_0^2(2T-t)) - (\sigma^2+\sigma_0^2(2T-t))^2\right)\\
        & =  \frac{\sigma^{-2}}{(\sigma^2+\sigma_0^2(2T-t))^2} \left(\sigma_0^2(-2\sigma^2t)+\sigma_0^4(2T^2-4Tt+t^2) - \sigma^4 \right)\\
        & = M'(t).
      \end{aligned}				
    \enqs
    We then verify the ODE for U by deriving the function $U$ w.r.t. $t$:
    \beqs
      \begin{aligned}		
        U'(t) &= (\sigma^2+\sigma_0^2T)\frac{\omega^2(2\sigma_0^2(T-t))}{2\left( (\sigma^2 + \sigma_0^2t)(\sigma^2 + \sigma_0^2(2T-t)) \right)^{\frac{3}{2}}} \frac{\sqrt{(\sigma^2 + \sigma_0^2t)(\sigma^2 + \sigma_0^2(2T-t))}}{\sigma^2+\sigma_0^2T} \\
        &= -\frac{\sigma_0^4(T-t)}{(\sigma^2 + \sigma_0^2t)(\sigma^2 + \sigma_0^2(2T-t))},
      \end{aligned}
    \enqs
    and from the right part of the ODE, we find the equality:
    \beqs
      \begin{aligned}	
        -\frac{\sigma^2 \sigma_0^4}{(\sigma^2+\sigma_0^2t)^2}M(t) & = -\frac{\sigma_0^4}{(\sigma^2+\sigma_0^2t)^2} \frac{\sigma^2 + \sigma_0^2t}{\sigma^2+\sigma_0^2(2T-t)}(T-t)\\
        &=-\frac{\sigma_0^4(T-t)}{(\sigma^2 + \sigma_0^2t)(\sigma^2 + \sigma_0^2(2T-t))}\\
        &=U'(t).
      \end{aligned}
    \enqs


\vspace{7mm}

\bibliographystyle{chicago}

\bibliography{Bib_Projet}{}


\end{document}